\documentclass[12pt,twoside,pointlessnumbers,smallheadings]{revtex4}

\usepackage{hangcaption,fancybox,feynmp}
\usepackage{indentfirst}
\usepackage{graphicx}
\usepackage{epsfig,subfigure,psfrag}
\usepackage{CJK}
\usepackage{mathrsfs}

\renewcommand{\baselinestretch}{1.2}

\begin{document}


\title{ Forward-backward asymmetry and differential cross section of top quark in
flavor violating $Z'$ model at  $\mathscr{O}(\alpha_s^2 \alpha_X)$ }
\author{Bo Xiao$^{1}$\footnote{E-mail:homenature@pku.edu.cn},
You-kai Wang $^{1}$\footnote{E-mail:wangyk@pku.edu.cn}, Shou-hua
Zhu$^{1,2}$\footnote{E-mail:aiwen\_fan@pku.edu.cn}}

\affiliation{ $ ^1$ Institute of Theoretical Physics $\&$ State Key
Laboratory of Nuclear Physics and Technology, Peking University,
Beijing 100871, China \\
$ ^2$ Center for High Energy Physics, Peking University, Beijing
100871, China }

\date{\today}
\maketitle

\begin{center}
\begin{minipage}{15cm}
{\small {\bf Abstract} \hskip 0.25cm

In this paper, forward-backward asymmetry and differential cross
section of top quark in flavor violating $Z'$ model up to
$\mathscr{O}(\alpha_s^2 \alpha_X)$ at Tevatron are calculated.  In
order to account for the top observed large forward-backward
asymmetry, the new coupling $g_X$ among $Z'$ and quarks will be not
much less than strong coupling constant $g_s$. After including the
new higher order correction, the differential cross section can fit
the data better than those only including the leading contributions
from $Z'$, while the forward-backward asymmetry is still in
agreement with the measurement.
 }

\end{minipage}
\end{center}

\renewcommand{\baselinestretch}{1.2}
\fontsize{12pt}{12pt}\selectfont

\section{Introduction\label{introduction}}

As the heaviest fermion in the standard model (SM), top quark is
thought to be closely related to the mechanism of electroweak
symmetry breaking and physics beyond the SM (BSM). In the last two
years, D0 and CDF Collaboration have measured the forward-backward
(F-B) asymmetry ($A_{FB}$) of top quark at the Tevatron
\cite{PhysRevLett.100.142002,Aaltonen:2008hc,Aaltonen:2009iz}. SM
predictions have been estimated in
Refs.~\cite{Kuhn:1998jr,Kuhn:1998kw,Antunano:2007da}. In the SM, the
asymmetry arises from the interference among virtual box and the
leading  diagrams for the process $q\bar q \rightarrow t \bar t$, as
well as the contributions from $q\bar q \rightarrow t \bar t g$. The
present experimental measurements and SM theoretical predictions are
listed in Table~\ref{Afb} in the lab ($p\bar{p}$) frame and the
center-of-mass (c.m.) frame of the top quark pair ($t\bar{t}$),
respectively. From the table we can see that the CDF measured
$A_{FB}^{p\bar{p}}$ is consistent with $A_{FB}^{t\bar{t}}$, if the
theoretically expected dilution of $30\%$ is included
\cite{Antunano:2007da}. However the SM predictions is significantly
smaller than the observations.

\begin{table}[htb]
\caption{\label{Afb}A collection of experimental and theoretical
results of $A_{FB}$ of top quark at the Tevatron
\cite{PhysRevLett.100.142002,Aaltonen:2008hc,Aaltonen:2009iz,Kuhn:1998jr,Kuhn:1998kw,Antunano:2007da}.
} \center \small{
\begin{tabular}{ccccc}\hline\hline
&D0 $(0.9 \mbox{fb}^{-1})$&CDF $(1.9 \mbox{fb}^{-1})$&CDF $(3.2 \mbox{fb}^{-1})$&SM Theory\\
\hline $A^{t\bar{t}}_{FB}$&
\begin{tabular}{c} $0.19\pm0.09\pm0.02$ (exclusive
4 jet)\\$0.12\pm0.08\pm0.01$ (inclusive 4 jet)\end{tabular} &$0.24\pm0.14$& $\cdots$ & 0.078(9)\\
 $A^{p\bar{p}}_{FB}$& $\cdots$ &
$0.17\pm0.08$&$0.193\pm0.065\pm0.024$&0.051(6)\\
\hline\hline
\end{tabular}
}
\end{table}

Recently some theoretical progress has been made both in the SM and
the BSM, in order to explain this novel signature. In the SM, soft
gluon resummation effects \cite{Almeida:2008ug} have been
scrutinized. However, the prediction involving resummation effects
does not change the asymmetry at $\mathscr{O}(\alpha_s^3)$ greatly
\cite{Almeida:2008ug}. Many BSM  models, for instance,
supersymmetry, extra dimension and left-right model have also been
considered
\cite{Frampton:2009rk,Shu:2009xf,Jung:2009jz,Cheung:2009ch,Cao:2010zb,
Djouadi:2009nb, Jung:2009pi,
Cao:2009uz,Barger:2010mw,Arhrib:2009hu}. New particles such as
exotic gluon $G'$, extra $W'$ or $Z'$ bosons and extra scalar $S$
are introduced. All these new models should produce the required
asymmetry while keep other observable qualities to be consistent
with measurements. Among which, the $t\bar{t}$ invariant mass
distribution is an important measurement to constrain the new
models. In order to distinguish different models, as depicted in the
Ref.~\cite{Cao:2010zb}, higher order effects in these new models are
important.

In this paper, we are interested in a BSM model named the flavor
violating $Z'$ model (FVZM) \cite{Jung:2009jz}. Observed asymmetry
 can be generated by introducing a right-handed coupling
 among the
$Z'$, the top and up quarks ${\cal{L}}\ni g_X
{Z_\mu}'\bar{u}\gamma^\mu\frac{1+\gamma^5}{2}t+\mbox{H.c.}$ A
detailed analysis based on leading order (LO) contributions has been
given in Ref.~\cite{Jung:2009jz}. For the suitable parameters, while
the asymmetry can be generated, the $t\bar{t}$ invariant mass
distribution does not fit the observation well. Therefore it is
quite interesting to analyze the asymmetry and $t\bar{t}$ invariant
mass distribution after including higher order effects.

The paper is organized as follows. In Sec. \ref{calculation}, Born,
virtual and real corrections are calculated analytically till to
$\mathscr{O}(\alpha_s^2 \alpha_X)$. In Sec. \ref{numerical},
numerical results for differential cross sections and
forward-backward asymmetry as a function of the top quark pair
invariant mass $M_{t\bar{t}}$ are presented and compared to the
experimental data. In Sec. \ref{summary}, we give a short
conclusions and discussions.

\section{Analytical calculation up to $\mathscr{O}(\alpha_s^2 \alpha_X)$ \label{calculation}}

In this section, we will present the analytical formula to calculate
the top forward-backward asymmetry, as well as the differential
cross section up to $\mathscr{O}(\alpha_s^2 \alpha_X)$. The
corresponding Feynman diagrams of subprocesses up to
$\mathscr{O}(\alpha_s^2 \alpha_X)$ in FVZM are depicted in Figs.
1-5. In order to account for the top large asymmetry, the new
coupling $g_X$ will be not much less than $g_s$. Thus the relevant
amplitude for $t\bar t$ final states can be written, in perturbation
series of couplings, as
\begin{eqnarray}
\mathscr{M}^{t\bar t} = f_s \alpha_s +f_X \alpha_X + f_s^1
\alpha_s^2+ f_{sX}^1 \alpha_s\alpha_X  + \cdots
\end{eqnarray}
with $\alpha_s=g_s^2/(4\pi)$, $\alpha_X=g_X^2/(4\pi)$, and $f$'s the
corresponding form factors. Squaring the amplitude we obtain
\begin{eqnarray}
|\mathscr{M}^{t\bar t}|^2 &=& |f_s|^2 \alpha^2_s+
 2\mathscr{R}\left( f_s^* f_X \right) \alpha_s \alpha_X + |f_X|^2 \alpha^2_X
 \nonumber \\
 && +
 2\mathscr{R}\left( f_s^* f_s^1\right) \alpha_s^3+ 2 \mathscr{R}\left(
 f_s^* f_{sX}^1+f_X^* f_s^1 \right)
\alpha^2_s\alpha_X +  \cdots \label{ttgeneral}
\end{eqnarray}

In order to cancel the infrared divergences, the corresponding gluon
radiation processes should be included. The amplitude for $t\bar t
g$ finals states can be written similarly as
\begin{eqnarray}
\mathscr{M}^{t\bar t g} = f^r_s \alpha_s \sqrt{\alpha_s} +f^r_X
\alpha_X \sqrt{\alpha_s} + \cdots
\end{eqnarray}
Squaring this amplitude we obtain
\begin{eqnarray}
|\mathscr{M}^{t\bar t g}|^2 &=&|f^r_s|^2 \alpha^3_s +
2\mathscr{R}\left(f^{r*}_s f^r_X\right)\alpha^2_s\alpha_X+ \cdots
\label{ttggeneral}
\end{eqnarray}

In the SM, the asymmetry arises from the $\mathscr{O}(\alpha_s^3)$
term. In the FVZM, new contributions till to
$\mathscr{O}(\alpha_X^2)$ are calculated in~\cite{Jung:2009jz}. In
this paper, the extra contributions at
$\mathscr{O}(\alpha_s^2\alpha_X)$ will be calculated. The SM $u\bar
u\to t\bar t$, $d\bar d\to t\bar t$ up to QCD NLO and $g g \to t\bar
t$ up to QCD LO contributions are recalculated though their
analytical expressions are not shown in this paper.

\subsection{Contributions up to $\mathscr{O}\left(\alpha^2_X\right)$}

\begin{figure}[htbp]
\centerline{\hbox{
\includegraphics[height=2.5cm,width=2.5cm]
{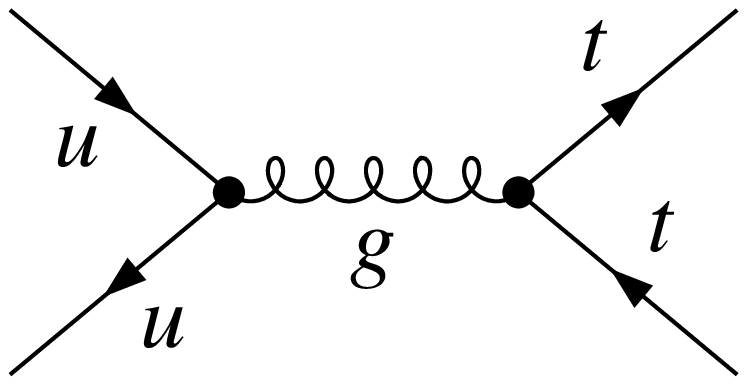}
\includegraphics[height=2.5cm,width=2.5cm]
{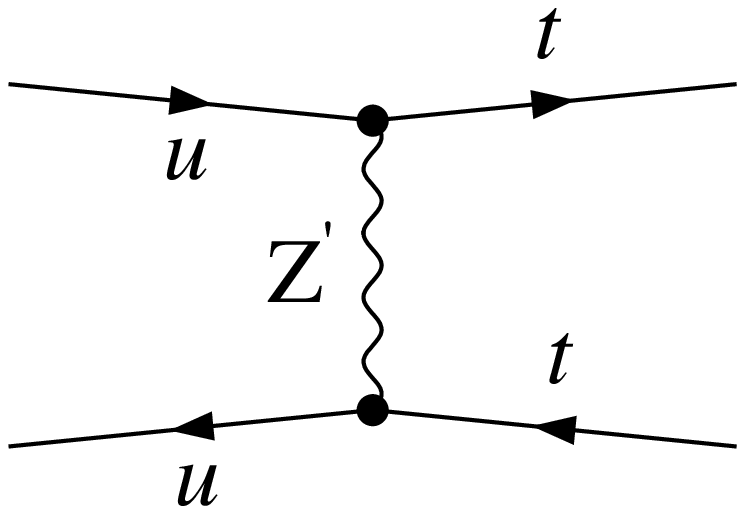} }} \caption{\label{tree}
Typical Feynman diagrams with contributions for form factors $f_s $
and $f_X$.}
\end{figure}

Typical Feynman diagrams, which contribute to the amplitude up to
$\mathscr{O}\left(\alpha^2_X\right)$, are shown in Fig.~\ref{tree}.

The form factors of $2\mathscr{R}\left( f_s^* f_X \right) $ and
$|f_X|^2$ with spin- and color-summed (same for the following form
factors) are given by
\begin{eqnarray}
2\mathscr{R}\left( f_s^* f_X \right)&=&\frac{64 \pi^2 C_A
C_F}{m_{Z'}^2 s \left(t-m_{Z'}^2\right)} \left[ m_t^6+\left(2
m_{Z'}^2 + s-2 t\right) m_t^4 \right. \nonumber
\\&& \left. +\left(t^2-2 m_{Z'}^2 (s+2
t)\right)
m_t^2+2 m_{Z'}^2 (s+t)^2\right]\label{RfsfX4}, \\
|f_X|^2&=&\frac{144 \pi^2 }{m_{Z'}^4
   \left(t-m_{Z'}^2\right)^2} \left[ m_t^8-2 t m_t^6+\left(4 m_{Z'}^4
   +4 s m_{Z'}^2+t^2\right) m_t^4 \right. \nonumber \\  && \left. -8
m_{Z'}^4 (s+t) m_t^2+4 m_{Z'}^4 (s+t)^2\right],
\end{eqnarray}
where $C_A=3, C_F=4/3$ and $s=(p_1+p_2)^2, t=(p_1-k_1)^2$ are the
Mandelstam variables.

\subsection{Contributions at $\mathscr{O}\left(\alpha^2_s \alpha_X\right)$}

The corresponding Feynman diagrams related to
$\mathscr{O}\left(\alpha^2_s \alpha_X\right)$ are shown in Figs.
2-5. In order to regulate the divergences, dimensional
regularization is adopted with $D=4-2\epsilon$. Infrared (IR) and
ultra violet (UV) divergences are represented by $1/\epsilon_{IR}$
and $1/\epsilon_{UV}$ respectively. The wave function
renormalization constants are determined by the on-mass-shell scheme
while $\overline{MS}$ scheme is chosen for the strong coupling
constants renormalization.  The calculations are carried out with
the help of FeynCalc~\cite{Mertig:1990an},
FormCalc~\cite{Hahn:1998yk} and QCDloop~\cite{Ellis:2007qk}. At
hadron collider, in order to eliminate the collinear singularity,
factorization should be carried out. In this paper $\overline{MS}$
factorization is adopted, as shown explicitly below.

\begin{figure}[htbp]
\centerline{\hbox{
\includegraphics[height=2.5cm,width=2.5cm]
{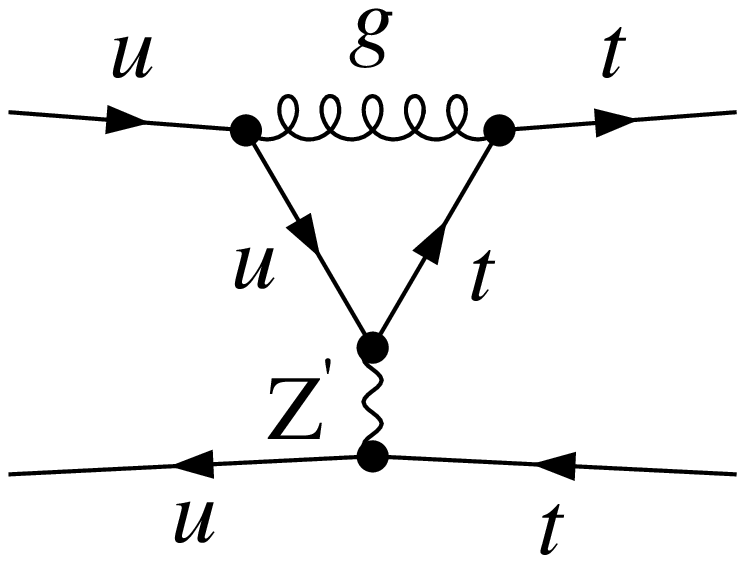}
\includegraphics[height=2.5cm,width=2.5cm]
{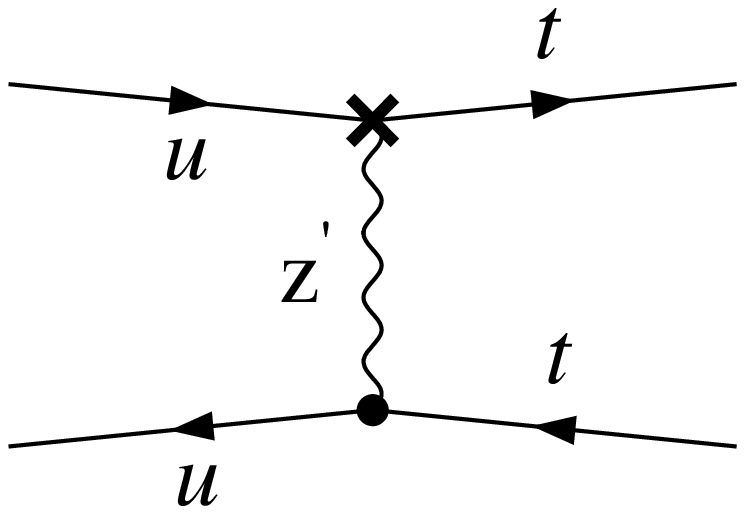}
\includegraphics[height=2.5cm,width=2.5cm]
{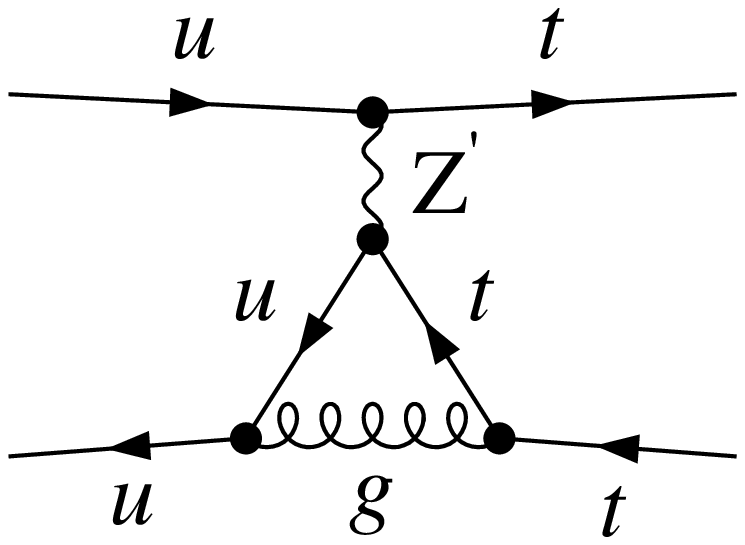}
\includegraphics[height=2.5cm,width=2.5cm]
{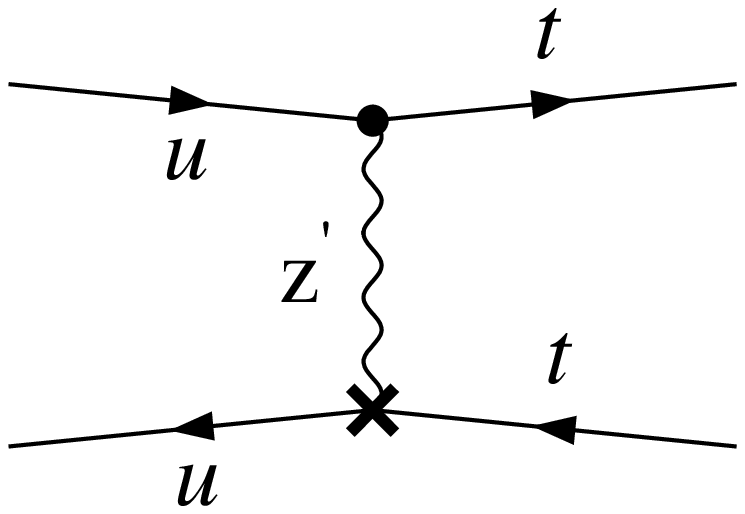} }}
\centerline{\hbox{
\includegraphics[height=2.5cm,width=2.5cm]
{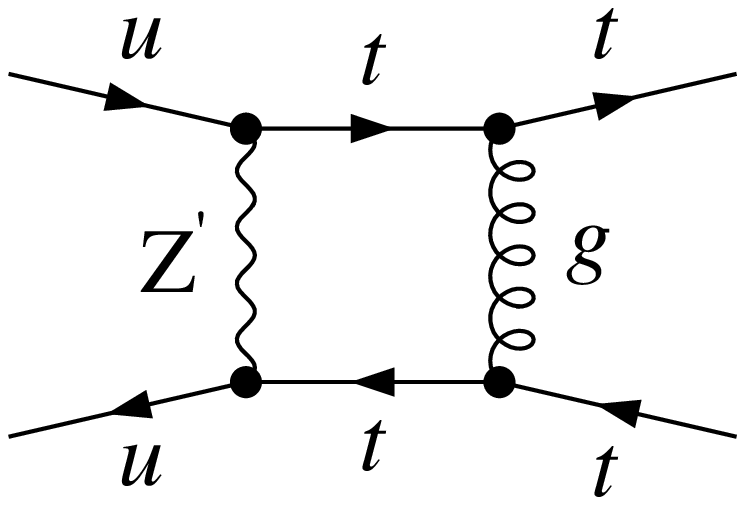}
\includegraphics[height=2.5cm,width=2.5cm]
{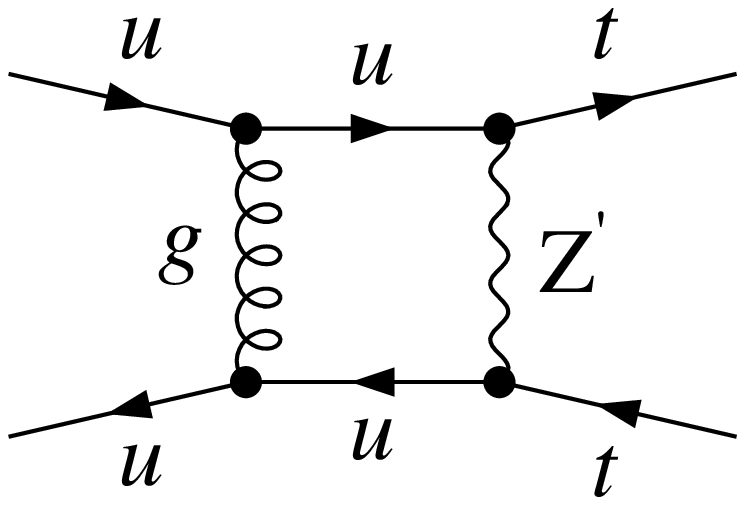}
\includegraphics[height=2.5cm,width=2.5cm]
{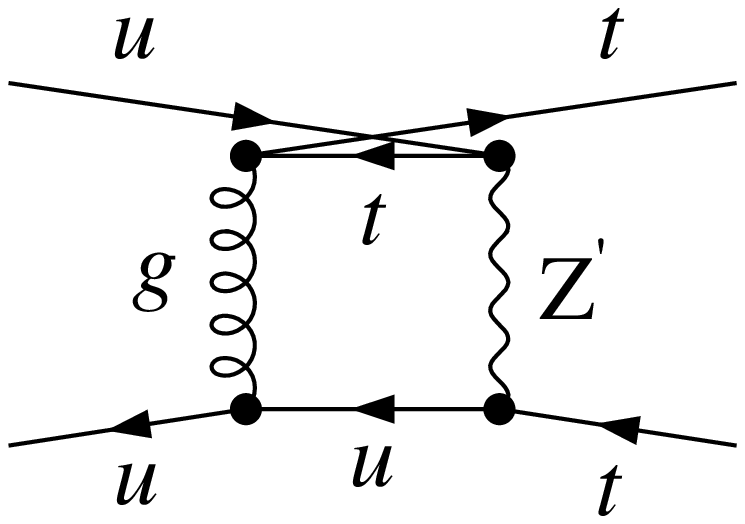}
\includegraphics[height=2.5cm,width=2.5cm]
{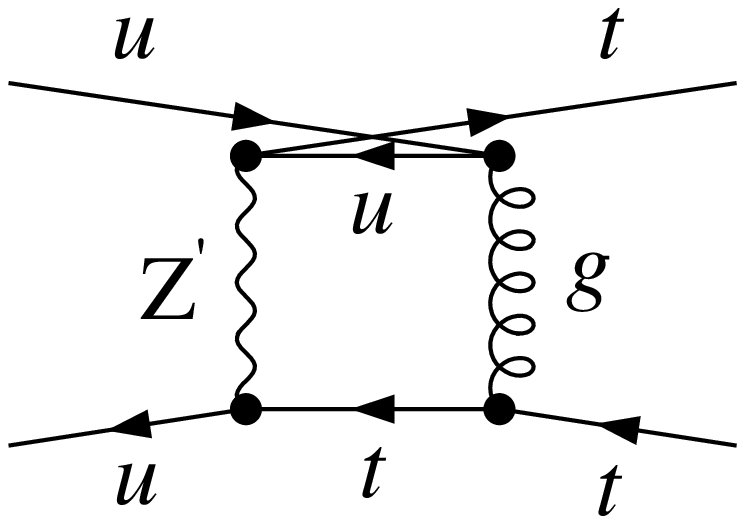} }}
\centerline{(a)} \centerline{\hbox{
\includegraphics[height=2.5cm,width=2.5cm]
{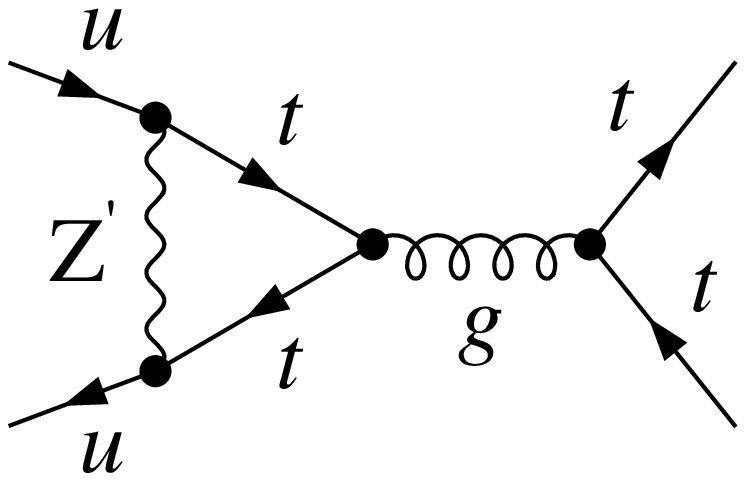}
\includegraphics[height=2.5cm,width=2.5cm]
{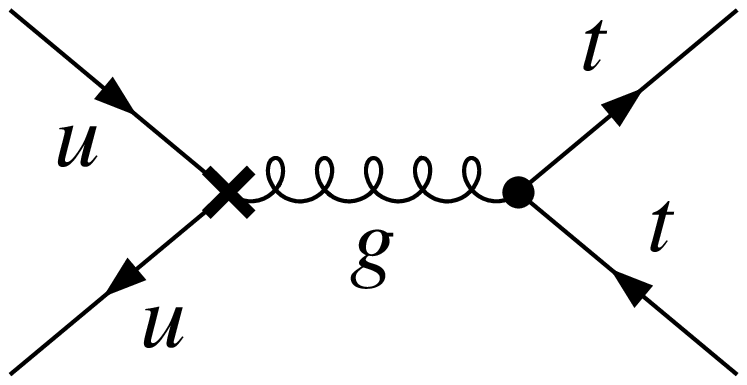}
\includegraphics[height=2.5cm,width=2.5cm]
{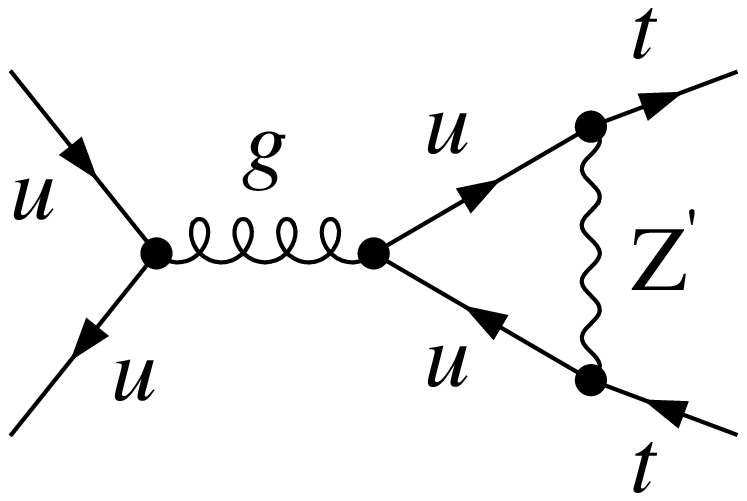}
\includegraphics[height=2.5cm,width=2.5cm]
{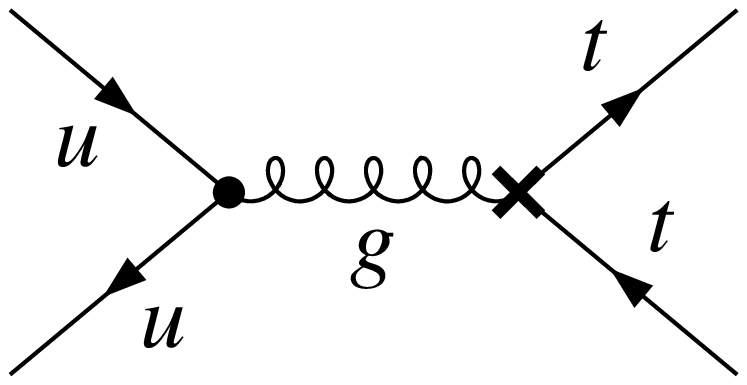} }}
\centerline{(b)} \caption{\label{zpvirtualdiagram} Typical Feynman
diagrams with contributions to form factors $f_{sX}^1 $.}
\end{figure}

\begin{figure}[htbp]
\centerline{\hbox{
\includegraphics[height=2.5cm,width=2.5cm]
{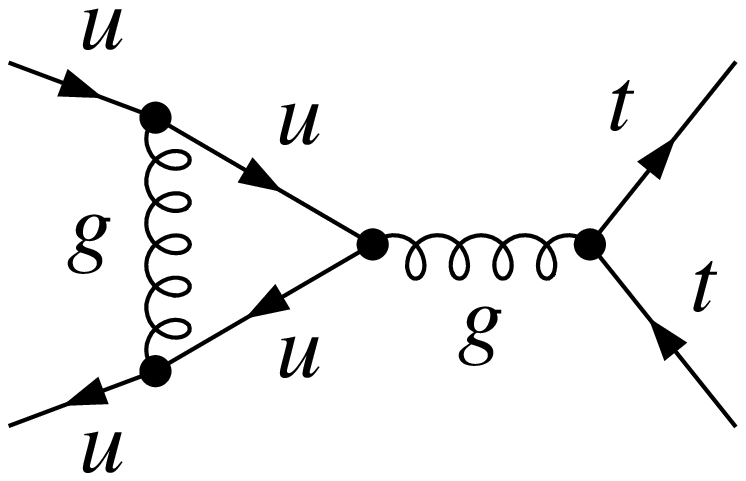}
\includegraphics[height=2.5cm,width=2.5cm]
{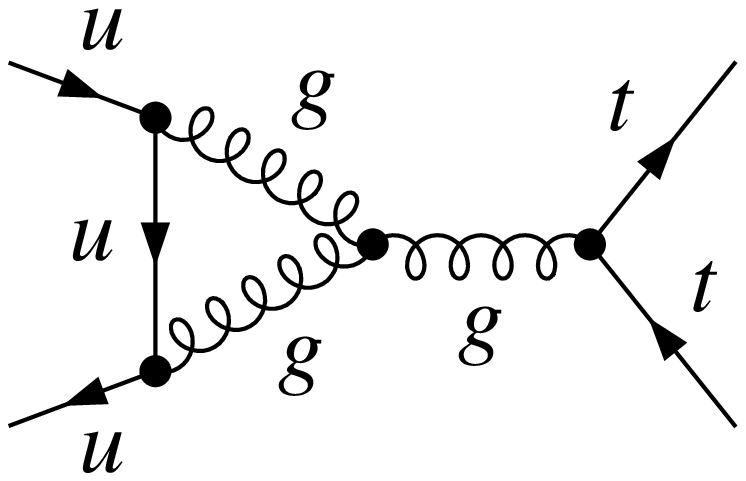}
\includegraphics[height=2.5cm,width=2.5cm]
{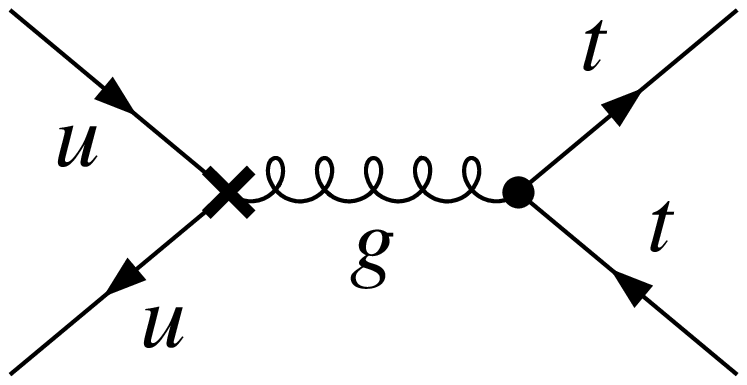}
\includegraphics[height=2.5cm,width=2.5cm]
{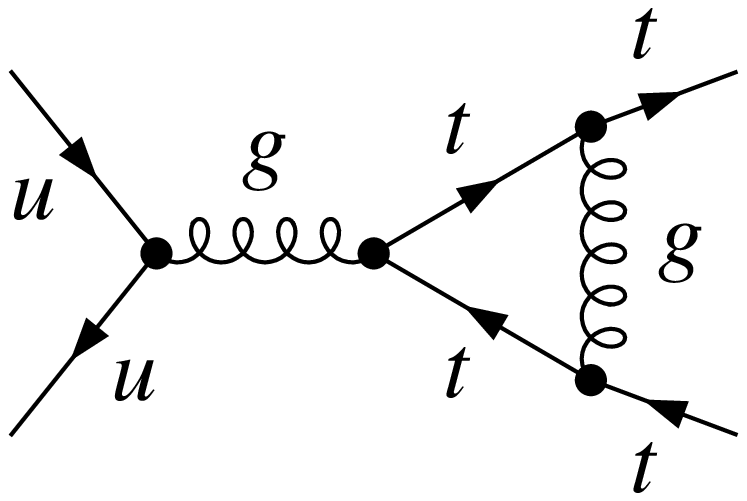}
\includegraphics[height=2.5cm,width=2.5cm]
{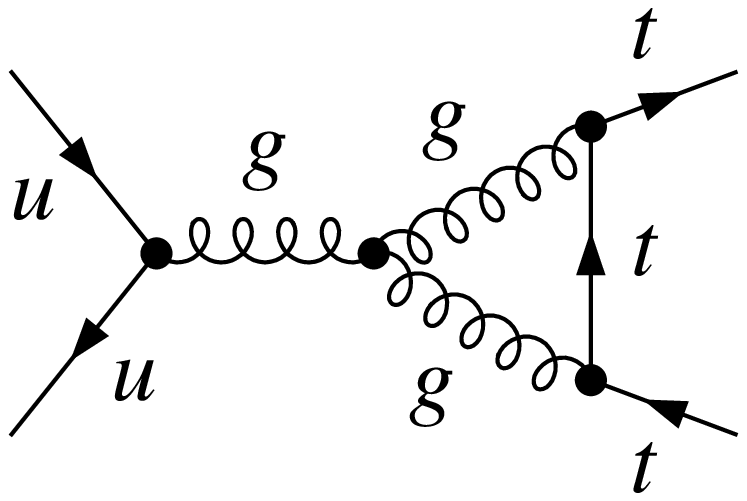}
\includegraphics[height=2.5cm,width=2.5cm]
{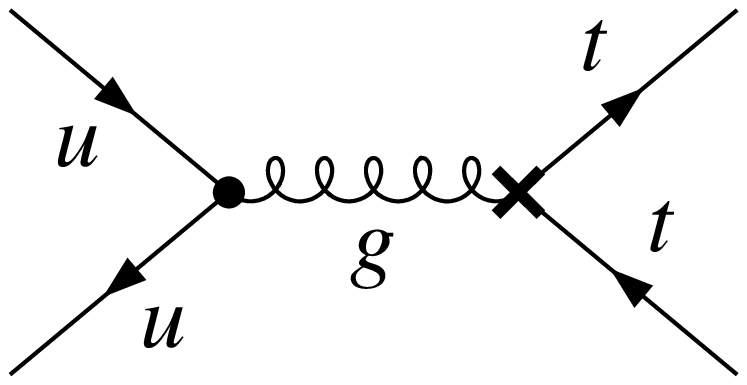}}}
\centerline{\hbox{
\includegraphics[height=2.5cm,width=2.5cm]
{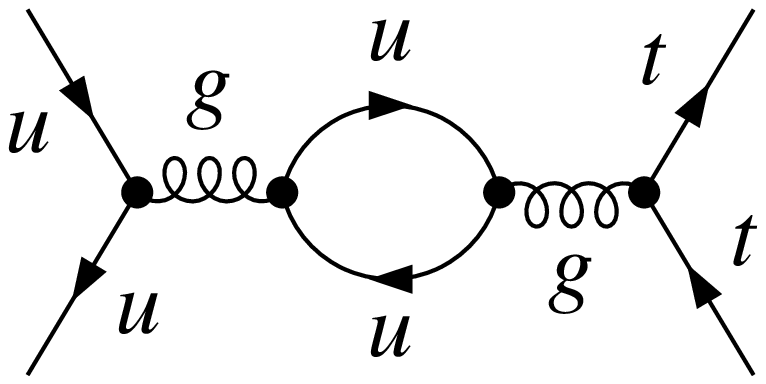}
\includegraphics[height=2.5cm,width=2.5cm]
{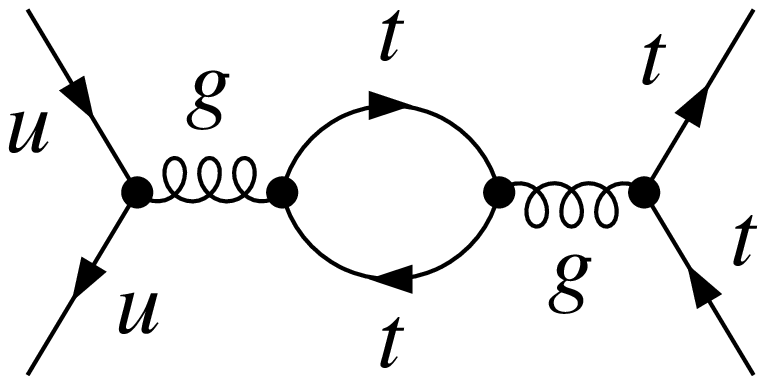}
\includegraphics[height=2.5cm,width=2.5cm]
{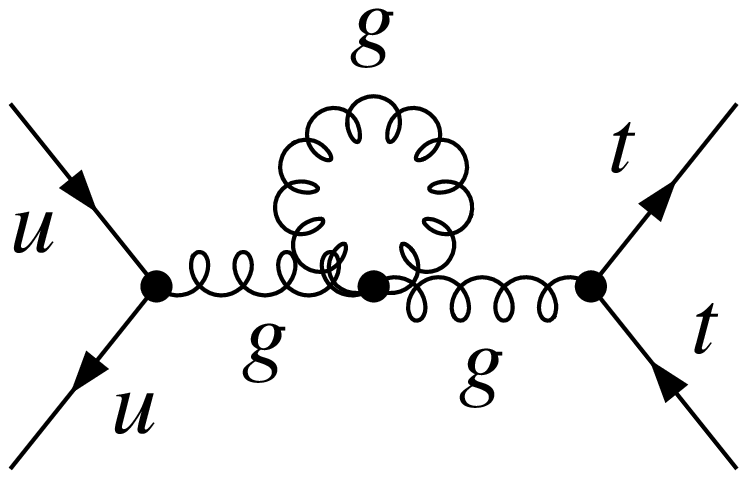}
\includegraphics[height=2.5cm,width=2.5cm]
{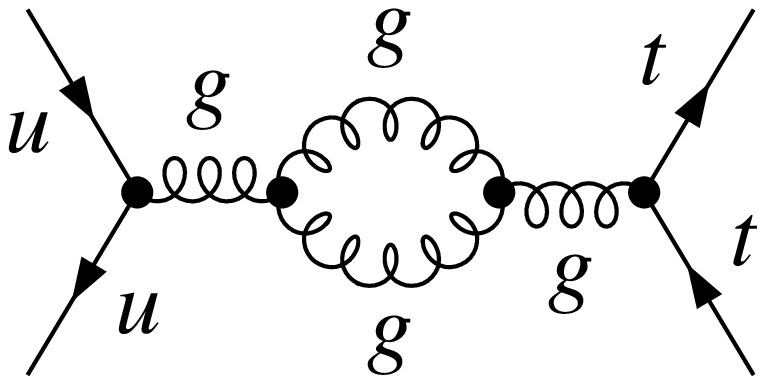}
\includegraphics[height=2.5cm,width=2.5cm]
{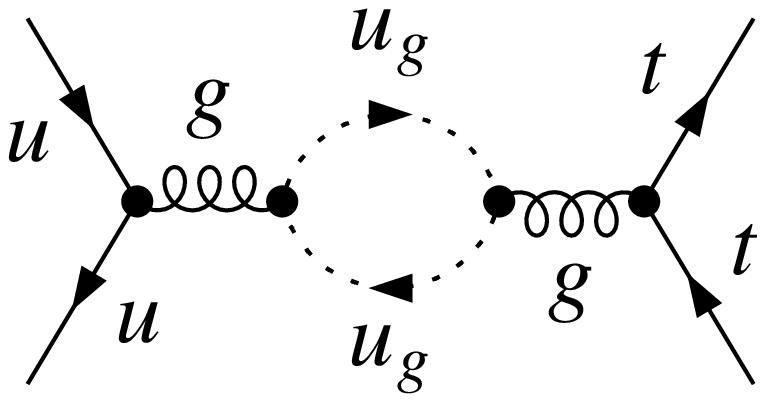}
\includegraphics[height=2.5cm,width=2.5cm]
{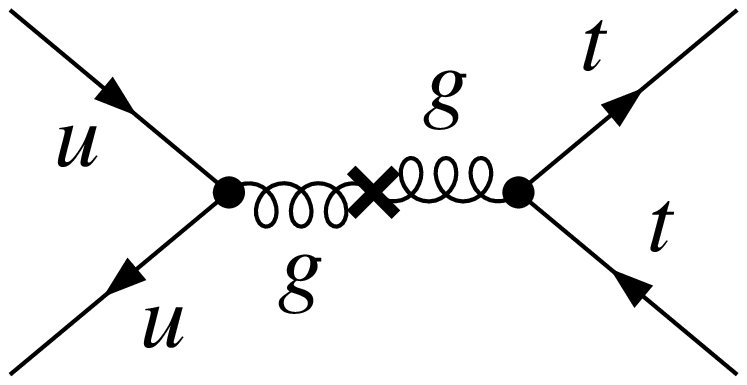}}}
\centerline{\hbox{
\includegraphics[height=2.5cm,width=2.5cm]
{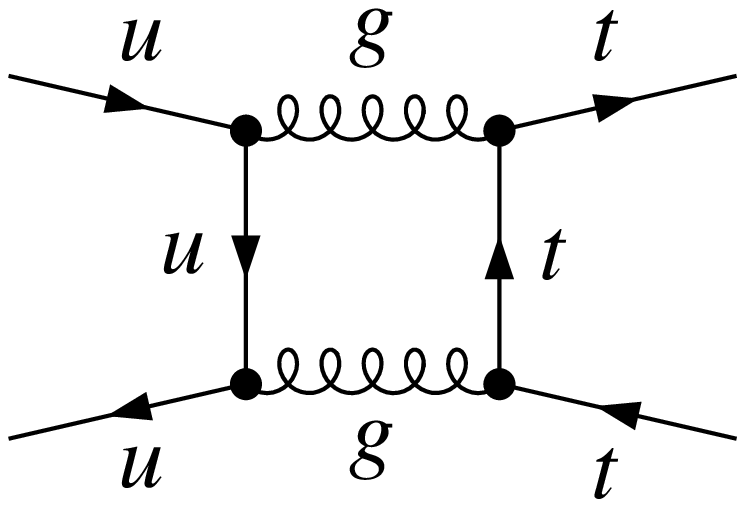}
\includegraphics[height=2.5cm,width=2.5cm]
{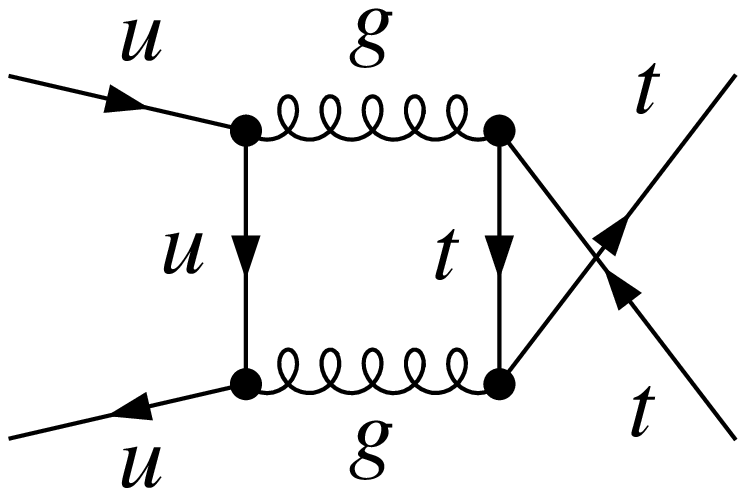} }}
\caption{\label{pureQCDvirtualdiagram} Typical Feynman diagrams with
contributions to form factors $f^1_{s} $.}
\end{figure}

\begin{figure}[htbp]
\centerline{\hbox{
\includegraphics[height=2.5cm,width=2.5cm]
{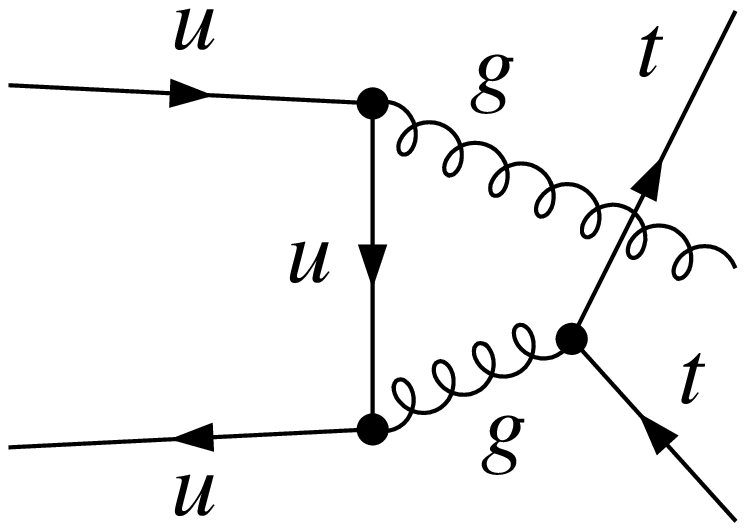}
\includegraphics[height=2.5cm,width=2.5cm]
{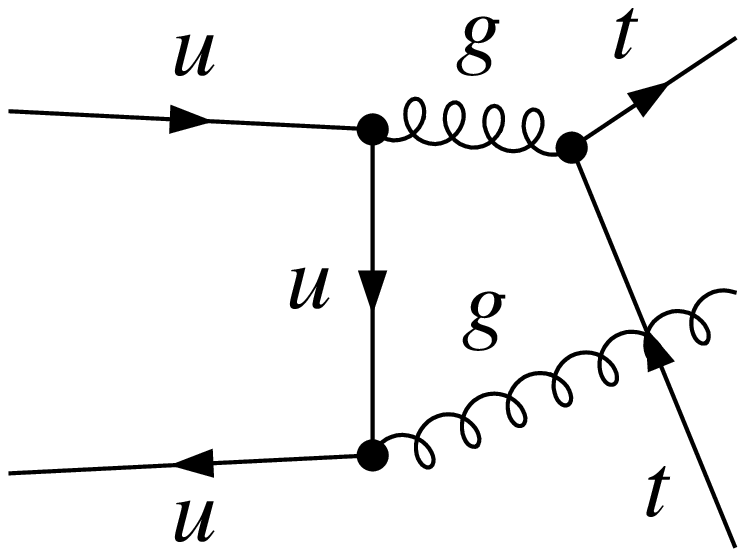}
\includegraphics[height=2.5cm,width=2.5cm]
{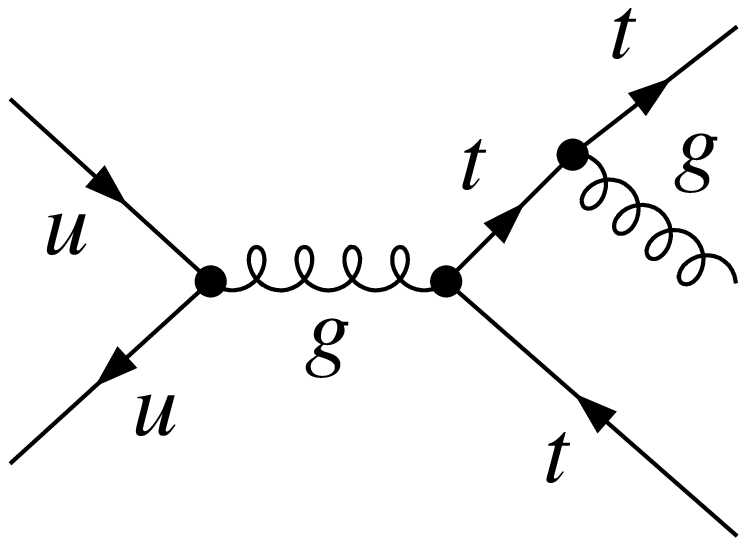}
\includegraphics[height=2.5cm,width=2.5cm]
{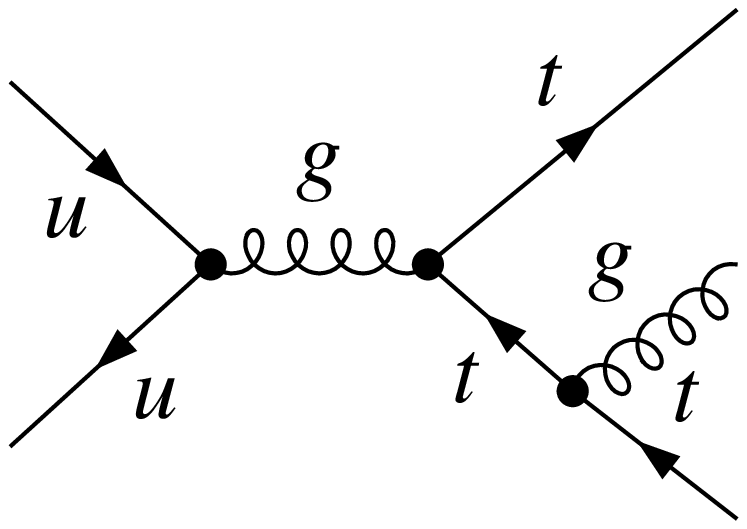}
\includegraphics[height=2.5cm,width=2.5cm]
{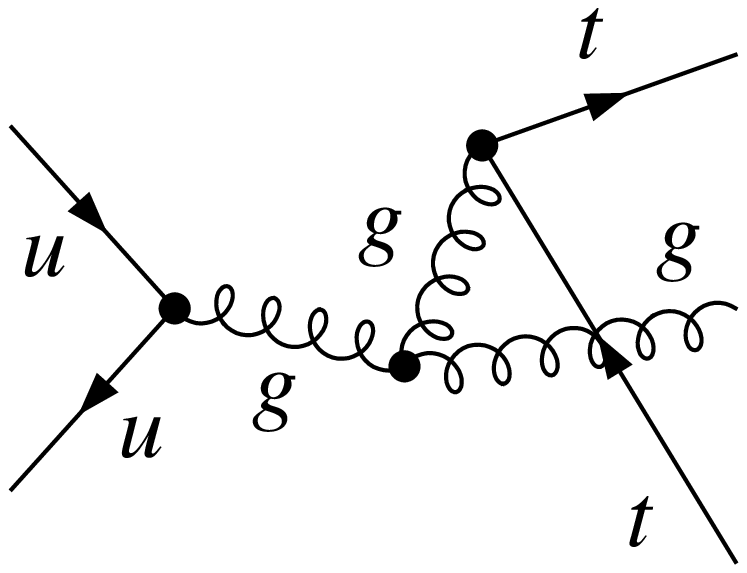}}}
\caption{\label{qcdrealgluonradiation}Typical Feynman diagrams with
contributions to form factors $f_{s}^r $.}
\end{figure}

\begin{figure}[htbp]
\centerline{\hbox{
\includegraphics[height=2.5cm,width=2.5cm]
{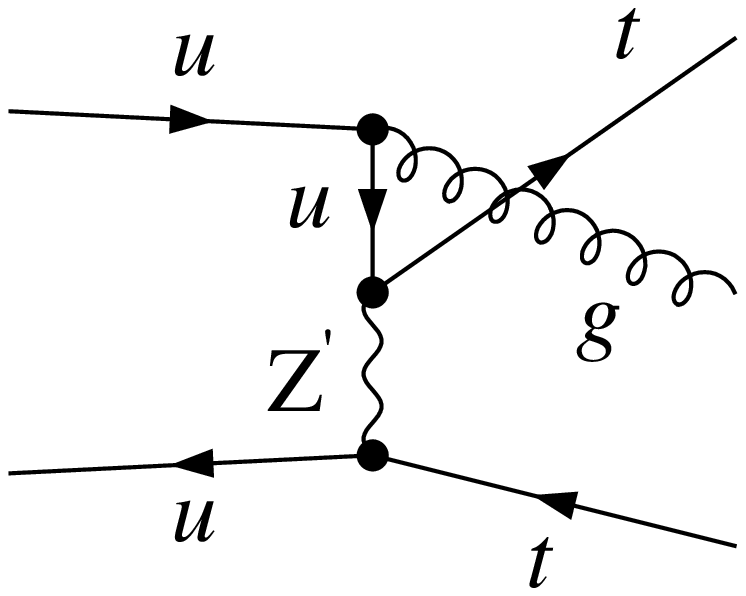}
\includegraphics[height=2.5cm,width=2.5cm]
{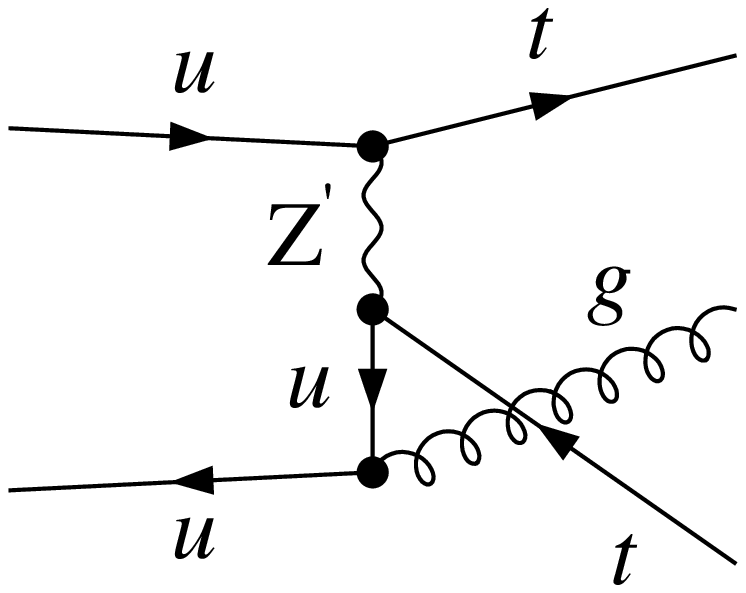}
\includegraphics[height=2.5cm,width=2.5cm]
{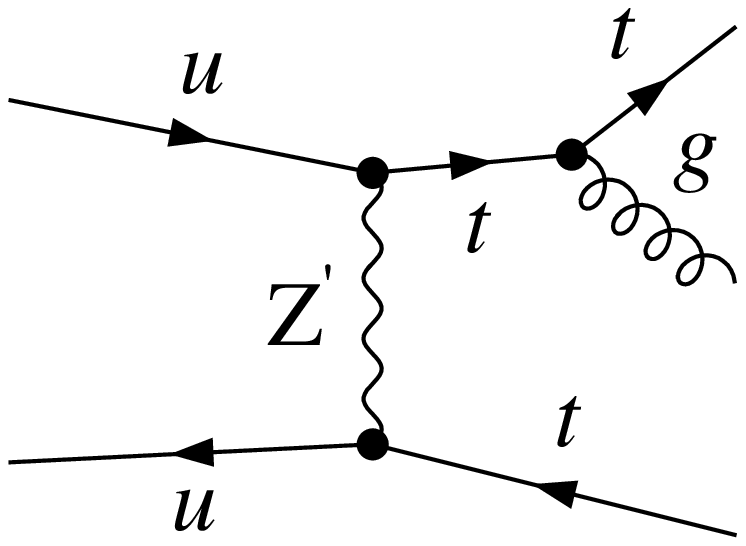}
\includegraphics[height=2.5cm,width=2.5cm]
{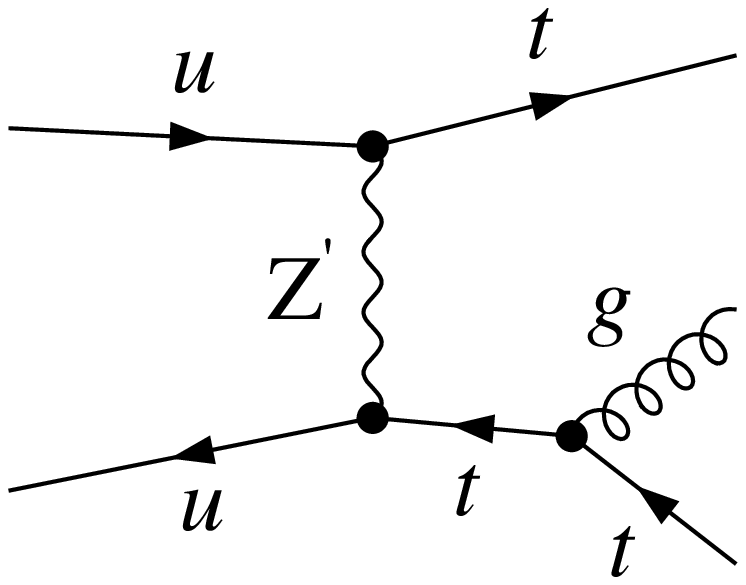}}}
\caption{\label{zprealgluonradiation}Typical Feynman diagrams with
contributions to form factors $f^r_{X}$.}
\end{figure}

Once the renormalization is carried out by adding appropriate
counterterms, the amplitude is free of UV divergences.  The counter
terms are given explicitly in Appendix A.   However there are still
IR divergences in $f_{sX}^1 $ and  $f_{s}^1 $. In order to eliminate
the remaining infrared divergence, the gluon emission processes
should be included as shown in Figs.~4 and 5. Two cutoff phase space
slicing method~\cite{Harris:2001sx} is applied for such processes.
First, small parameter $\delta_s$ is introduced to separate the
final phase space into soft part and hard part. Second, small
parameter $\delta_c$ is introduced to divide the hard part into a
hard collinear part and a hard non-collinear part. These three parts
are calculated separately. Their summation should be independent of
the two small parameters $\delta_s$ and $\delta_c$.

The remaining IR divergences appear in $2\mathscr{R}(f_s^*
f_{sX}^1+f_X^* f_s^1)\alpha_s^2\alpha_X$ [c.f. Eq.
[\ref{ttgeneral}]] can be expressed as
\begin{equation}
2\mathscr{R}(f_s^* f_{sX}^1+f_X^* f_s^1)\alpha_s^2\alpha_X= V_f +
V_1 \frac{1}{\epsilon_{IR}}+V_2 \frac{1}{\epsilon_{IR}^2},
\label{virtual}
\end{equation}
where $V_f$ indicates the finite part and IR coefficients $V_1, V_2$
are
\begin{equation}
\begin{array}{rl}
V_1=& [2\mathscr{R}(f_s^* f_X)'\alpha_s\alpha_X]
\frac{-2\alpha_s}{3\pi C_A
C_F}\frac{(4\pi)^\epsilon}{\Gamma(1-\epsilon)}
\left\{\frac{(-2m_t^2+s)}{s\beta}\log(\frac{1+\beta}{1-\beta})\right.\\\\
&\left.+9\log(\frac{m_t^2}{s})
+10\log(\frac{s}{\mu^2})-16\log(\frac{(m_t^2-t)}{\mu^2})-2\log(\frac{(-m_t^2+s+t)}{\mu^2})\right\}
\\\\
&-\frac{16(4\pi)^3\alpha_s^2\alpha_X}{3\pi^2}\frac{(4\pi)^\epsilon}{\Gamma(1-\epsilon)}
\frac{3m^4+3s^2+8st+3t^2-m^2(5s+6t)}{s(t-m_{Z'}^2)},\\\\

 V_2=&  [2\mathscr{R}(f_s^* f_X)'\alpha_s\alpha_X]\frac{-16\alpha_s
   }{ 3\pi C_A C_F }\frac{(4\pi)^\epsilon}{\Gamma(1-\epsilon)},
\end{array}
\end{equation}
where $\beta=\sqrt{1-4m_t^2/{s}}$  and $\mu$ is an energy scale
introduced in dimensional regularization. Here  \begin{equation}
2\mathscr{R}(f_s^* f_X)'=128\pi^2 C_A C_F
\frac{m_t^4+(s+t)^2-m_t^2(s+2t)}{s(t-m_{Z'}^2)}
\end{equation} is different from
$2\mathscr{R}(f_s^* f_X)$ in Eq.~[\ref{RfsfX4}] as Goldstone boson
contribution is ignored.

The IR poles in Eq.~[\ref{virtual}] have two physical origins: soft
\& collinear divergences. The double pole $1/\epsilon_{IR}^2$
indicates an overlap between soft and collinear divergences, and the
divergences can be eliminated by including contributions from the
soft region for gluon emission processes and parton distribution
function (PDF) redefinition \cite{Harris:2001sx}. Gluon emission
process in soft region can be calculated by the eikonal
approximation method (the details are given in Appendix B), which
can be expressed as
\begin{equation}
\left[2\mathscr{R}\left(f^{r*}_s
f^r_X\right)\alpha^2_s\alpha_X\right]_{\text{Soft}}=S_f
+R_1\frac{1}{\epsilon_{IR}}+R_2\frac{1}{\epsilon_{IR}^2},
\label{realsoft}
\end{equation}
where $S_f$ indicates the finite part and IR coefficients $R_1, R_2$
are
\begin{equation}
\begin{array}{rl}
R_1=&[2\mathscr{R}(f_s^* f_X)^D\alpha_s\alpha_X]\frac{\alpha_s}{3
\pi C_A C_F}
\frac{(4\pi)^\epsilon}{\Gamma(1-\epsilon)}\\\\
&\left\{\frac{1+\beta^2}{\beta}\log(\frac{1+\beta}{1-\beta})+16
+18\log(\frac{m_t^2}{s})
-32\log\delta_s\right.\\\\
&\left.+20\log(\frac{s}{\mu^2}) -32\log(\frac{(m_t^2-t)}{\mu^2})
-4\log(\frac{(s+t-m_t^2)}{\mu^2})\right\},\\\\
R_2=&[2\mathscr{R}(f_s^* f_X)^D\alpha_s\alpha_X] \frac{16\alpha_s
    }{ 3\pi C_A C_F}\frac{(4
\pi)^\epsilon}{\Gamma(1-\epsilon)},
\end{array}
\end{equation}
where $2\mathscr{R}(f_s^* f_X)^D$ is a D-dimension version of
$2\mathscr{R}(f_s^* f_X)'$,
\begin{equation}
2\mathscr{R}(f_s^* f_X)^D=128\pi^2 C_AC_F
\frac{(1-\epsilon_{IR})[m_t^4+(s+t)^2-m_t^2(s+2t)+s(t-m_t^2)\epsilon_{IR}]}{s(t-m_{Z'}^2)}
\end{equation}
PDF redefinition in soft region can be written as
\cite{Harris:2001sx},
\begin{equation}
[2 \mathscr{R}(f_s^* f_X)^D\alpha_s\alpha_X]\frac{\alpha_s}{4\pi}
\frac{(4\pi)^\epsilon}{\Gamma(1-\epsilon)} (3 C_F+4 C_F\log
\delta_s) (\log(\frac{\mu_f^2}{\mu^2})-\frac{1}{\epsilon_{IR}}).
\label{pdfcollinear}
\end{equation}
IR cancelation is realized by adding Eqs.~[\ref{virtual}] and
[\ref{realsoft}],
 and subtract 2 times of expression in Eq.~[\ref{pdfcollinear}].

For the hard collinear part for gluon emission processes, there are
collinear divergences which can be eliminated by PDF redefinition in
this region. The remaining finite part is in the form of a
convolution integral with a splitting function. The detail of this
procedure is described in Ref.~\cite{Harris:2001sx}.

The remaining hard noncollinear part of gluon emission processes is
finite and the integration is performed with a standard three body
Monte Carlo code.

During the calculation, there are three scales in the hadron level
cross section. $\mu$ is an energy scale introduced in dimensional
regularization. $\mu_r$ is a renormalization scale introduced in the
$\overline{MS}$ renormalization of the coupling constant. $\mu_f$ is
a factorization scale introduced in $\overline{MS}$ factorization.
The $\mu$ always comes with the divergences and is canceled
completely when the corresponding poles are subtracted. The
dependence of our final results on $\mu_r$ \& $\mu_f$ will be shown
in the numerical results.

The independence of the total cross section with $\delta_s$ and
$\delta_c$ has been checked in the situation $\delta_c<<\delta_s$,
as suggested by the two cut off phase space slicing method in
Ref.~\cite{Harris:2001sx}.

\section{Numerical Results\label{numerical}}

In this section, we will present the numerical results  and compare
them with the experimental measurements. We choose cteq6l for
leading order calculation and cteq6m for higher order calculations.
The scales $\mu_r$ and $\mu_f$ are set to be equal and
$\alpha_S(m_Z)=0.118$.

\begin{figure}[htbp]
\begin{center}
\includegraphics[width=0.9\textwidth]
{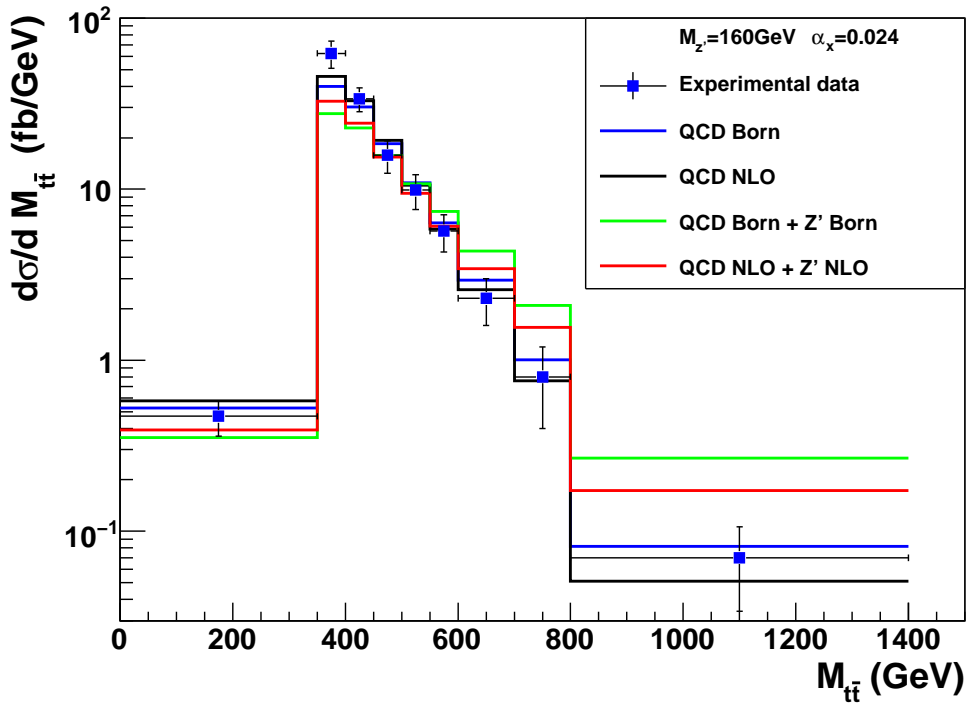}
 \caption{\label{bestpoint}
Differential cross sections as a function of $M_{t\bar t}$ with
$\mu_r=\mu_f=m_t$. Here ``QCD Born" and ``QCD NLO" represent the
results in the SM at leading order and next-to-leading order in QCD.
``QCD Born + $Z'$ Born" and ``QCD NLO + $Z'$  NLO" stand for the
predictions in FVZM up to $\mathscr{O}\left(\alpha^2_X\right)$ and
$\mathscr{O}\left (\alpha^2_s\alpha_X\right)$ respectively [c.f.
Eqs. \ref{ttgeneral} and \ref{ttggeneral} ].}
\end{center}
\end{figure}

Differential cross sections as a function of $M_{t\bar{t}}$ are
shown in Fig.~\ref{bestpoint}. Histograms are drawn here in order to
compare conveniently with the experimental measurements
\cite{Aaltonen:2009iz}. The parameters in the FVZM are taken to be
$\alpha_X=0.024$, $M_{Z'}=160\mbox{GeV}$ which is the best
point~\cite{Jung:2009jz} to account for the top asymmetry. From the
figure it is obvious that the NLO QCD prediction is in good
agreement with the data except the bin around 400 GeV. It should be
noted that even the multiple soft gluon radiation effects are
included, the discrepancy remains (c.f. Ref.~\cite{Almeida:2008ug}).
However the top quark asymmetry at NLO QCD is much less than the
measurement. After including the contributions from the leading
diagrams from $Z'$, for the favorable parameters, the top quark
asymmetry can be generated. However the differential cross section
does not agree with measurement well. From the figure, we can see
the prediction is lower than measurement for small $M_{t\bar t}$
region while higher for large $M_{t\bar t}$. After including the
higher order effects, the differential distribution will be better
while the top asymmetry can be generated. These behavior can be
understood as follows. In the vicinity of $t\bar t$ production
threshold region, the significant contributions comes from the
interference among QCD and extra $Z'$ Feynman diagrams.
 Such contributions will decrease the cross section.
 In the higher $M_{t\bar t}$ region, the square of $Z'$ diagrams become significant and they will
uplift the cross section. The deviation from NLO QCD prediction will
be soften after including the higher order contributions.

\begin{figure}[htbp]
\centerline{\hbox{
\includegraphics[width=0.5\textwidth]
{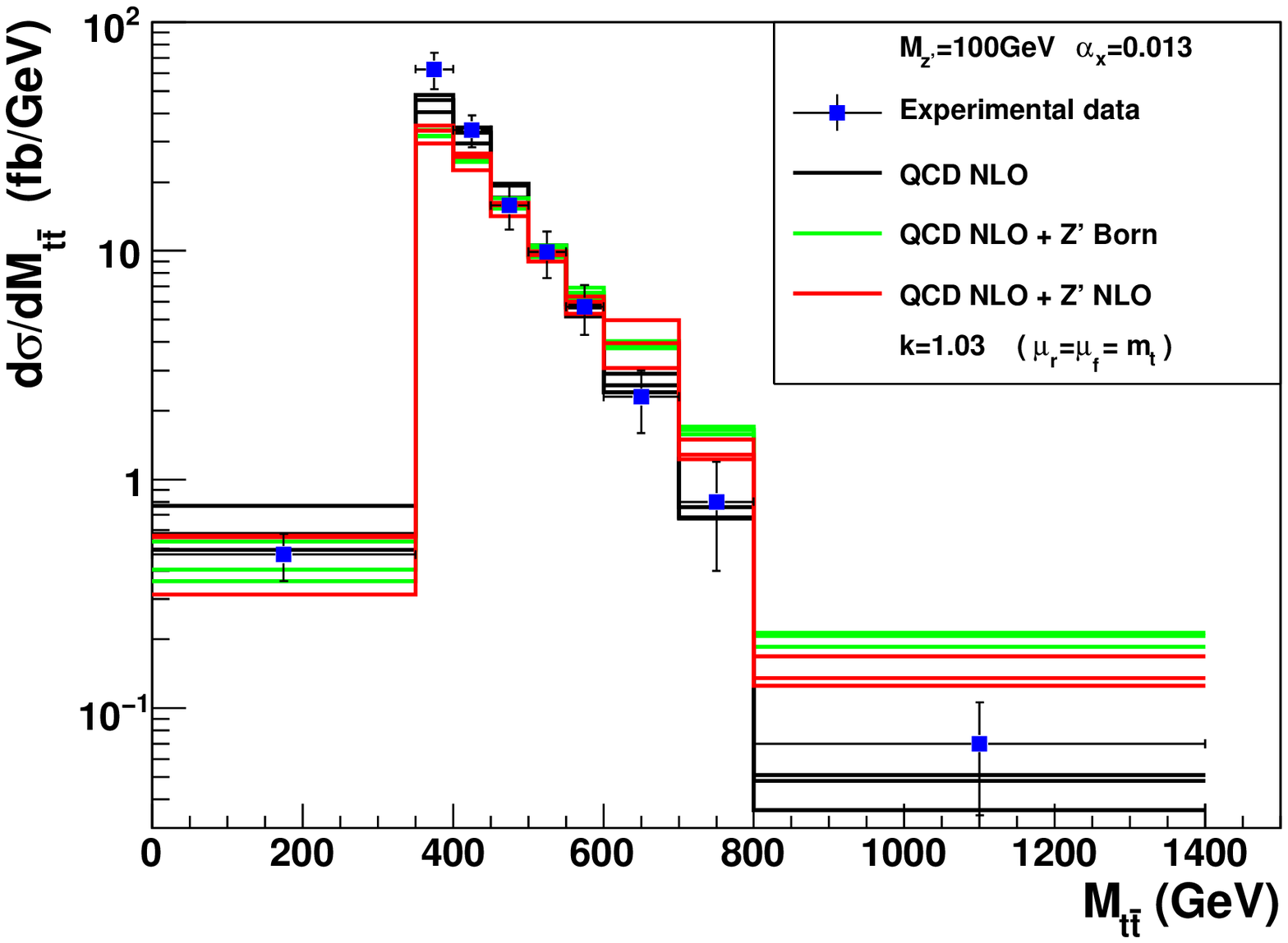}
\includegraphics[width=0.5\textwidth]
{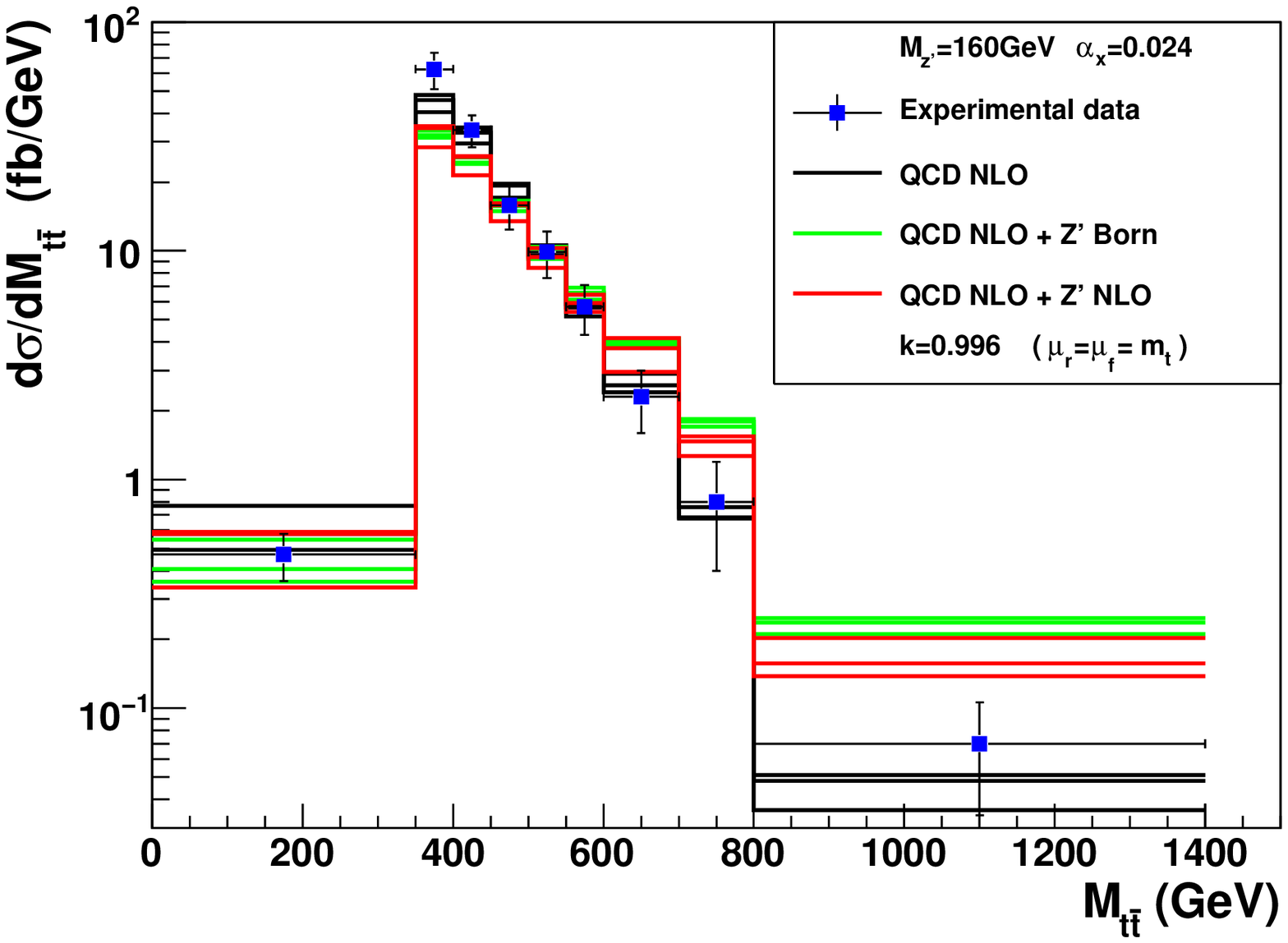}}} \centerline{\hbox{
\includegraphics[width=0.5\textwidth]
{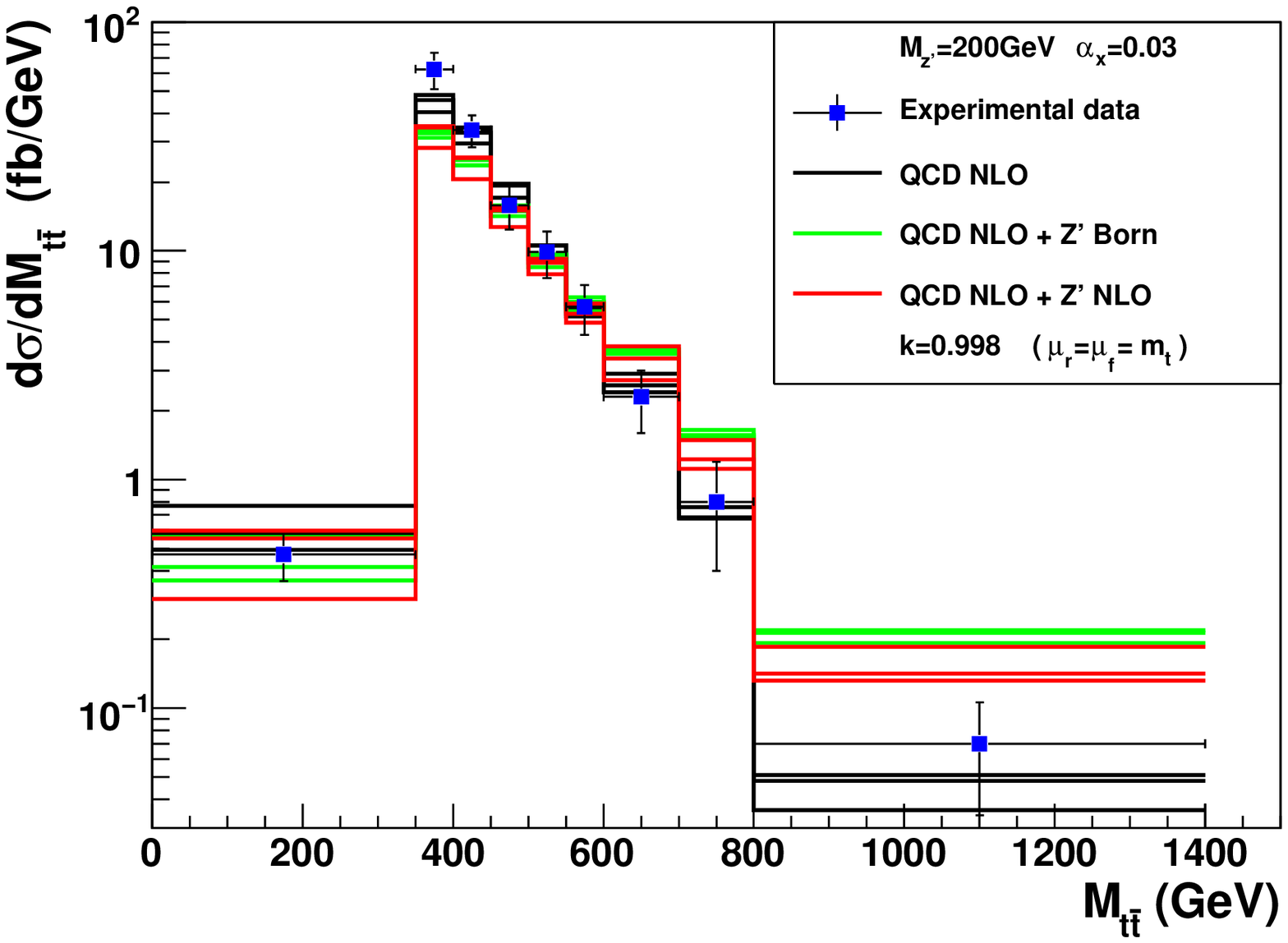}
\includegraphics[width=0.5\textwidth]
{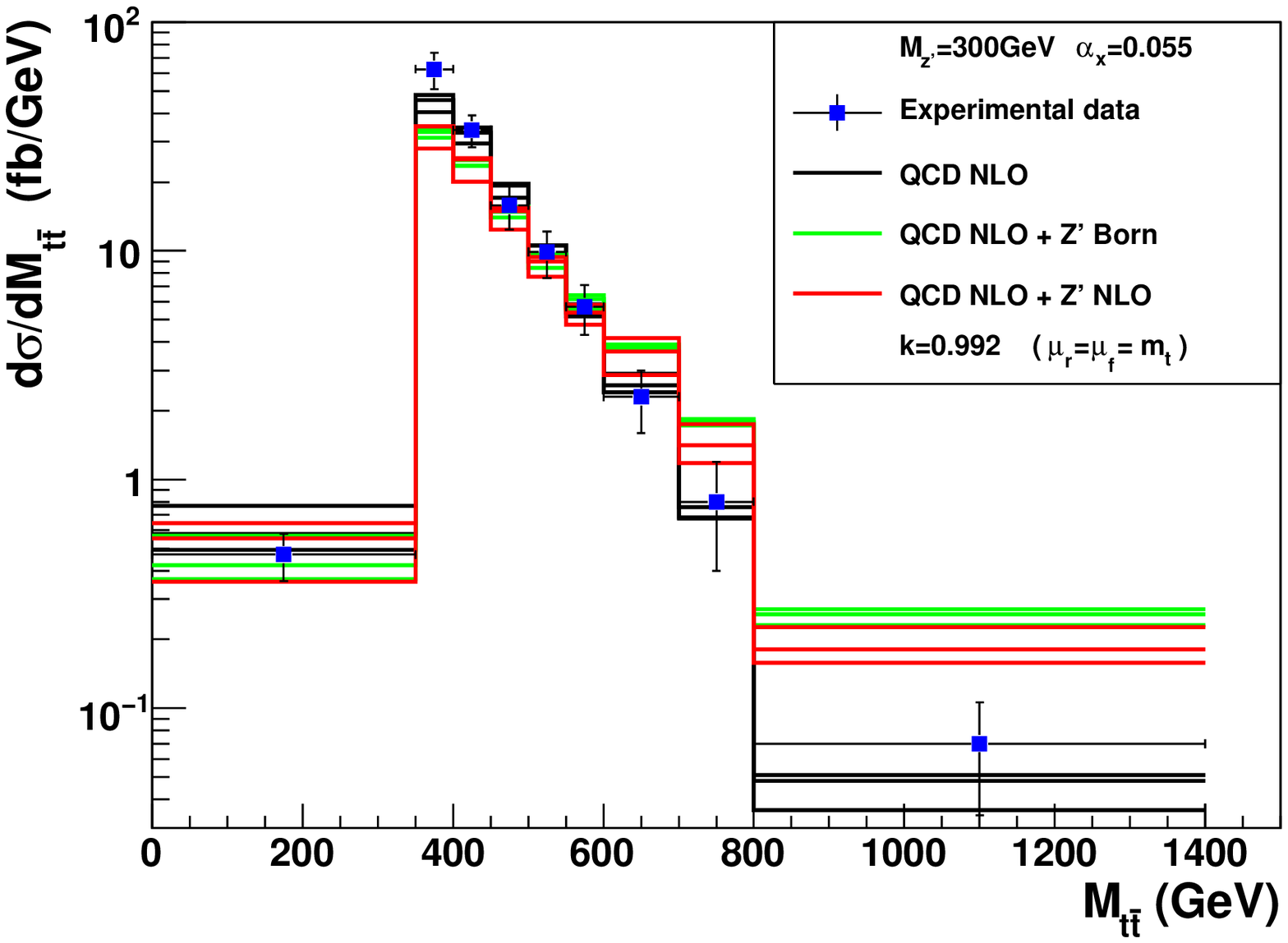}}} \caption{\label{4benchmarkpoints}
Differential cross sections $d\sigma/dM_{t\bar{t}}$  as a function
of  $M_{t\bar t}$ with $\mu_r=\mu_f=0.5m_t, m_t, 2m_t$,
respectively, for 4 sets of parameters. Other conventions are the
same with Fig.~\ref{bestpoint}.
 The factor $k\equiv\sigma^{NLO}/\sigma^{LO}$ is also indicated.}
\end{figure}

In Fig.~\ref{4benchmarkpoints} we show the differential cross
sections $d\sigma/dM_{t\bar{t}}$ with $\mu_r=\mu_f=0.5m_t, m_t,
2m_t$ respectively,
 for four sets of typical parameters, namely different $M_{Z'}$ and $\alpha_X$.
From the histograms, we can see that the above-mentioned improvement
after including the higher order effects is universal. We calculated
the k factor, which is defined as $k\equiv\sigma^{NLO}/\sigma^{LO}$.
Here $\sigma^{LO}$ and $\sigma^{NLO}$ are the cross sections up to
$\mathscr{O}(\alpha_X^2)$ and $\mathscr{O}(\alpha_S^2\alpha_X)$
respectively. For four sets of parameters, the k-factor is equal to
1.03, 0.996, 0.998, 0.992, respectively, for $\mu_r=\mu_f =m_t$. It
is obvious that the NLO contributions mainly change the shape of
distribution. As for the scale dependence, the results of ``QCD NLO"
and ``QCD NLO+$Z'$ NLO" are about the same size. ``QCD NLO+ $Z'$
Born" is significantly smaller than them. ``QCD NLO + $Z'$ Born"
scale dependence is small because QCD NLO and $Z'$ Born have
opposite $\mu_r/\mu_f$ dependence.

\begin{figure}[htbp]
\centerline{\hbox{
\includegraphics[width=0.8\textwidth]
{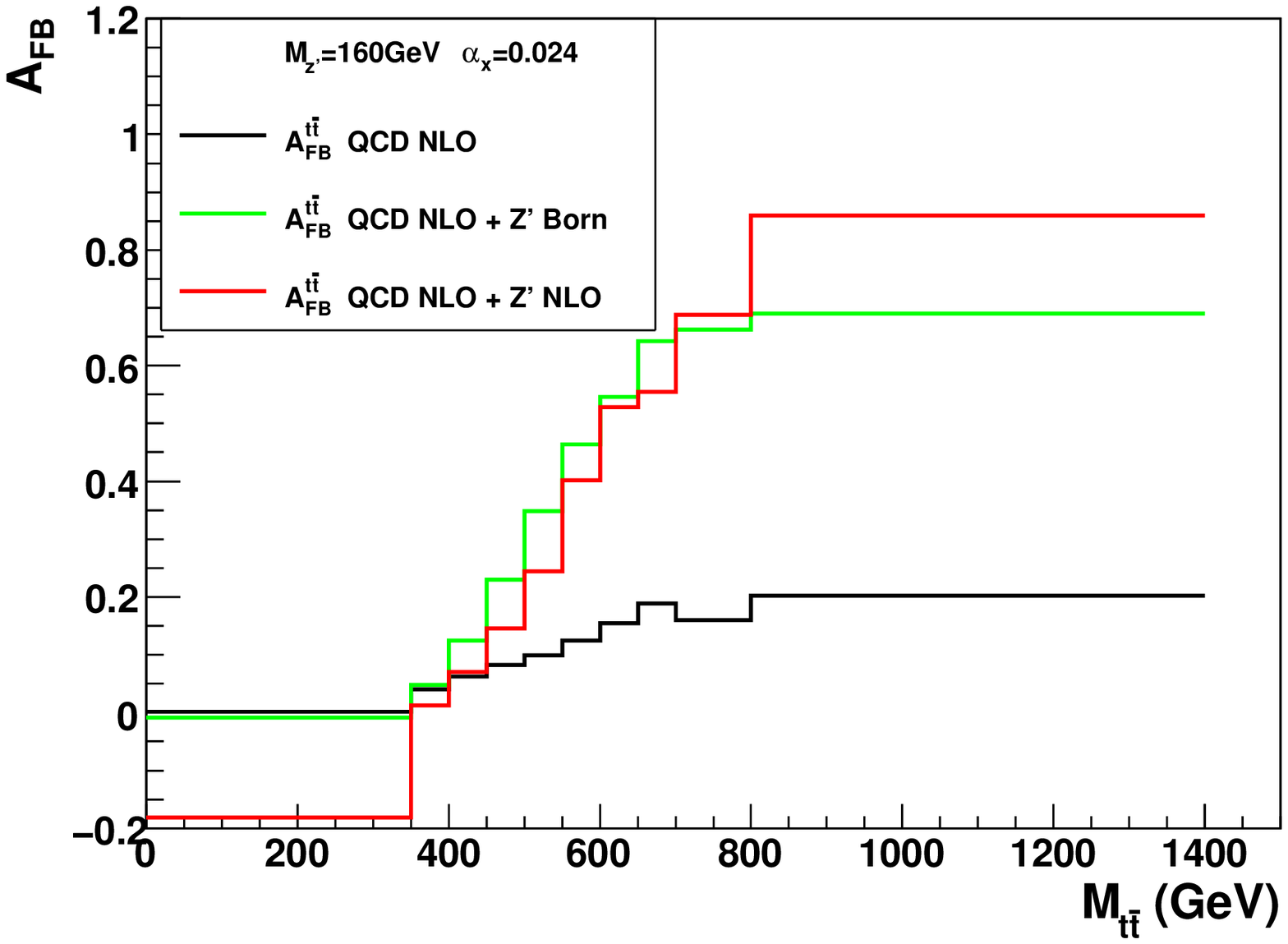}}} \centerline{\hbox{
\includegraphics[width=0.5\textwidth]
{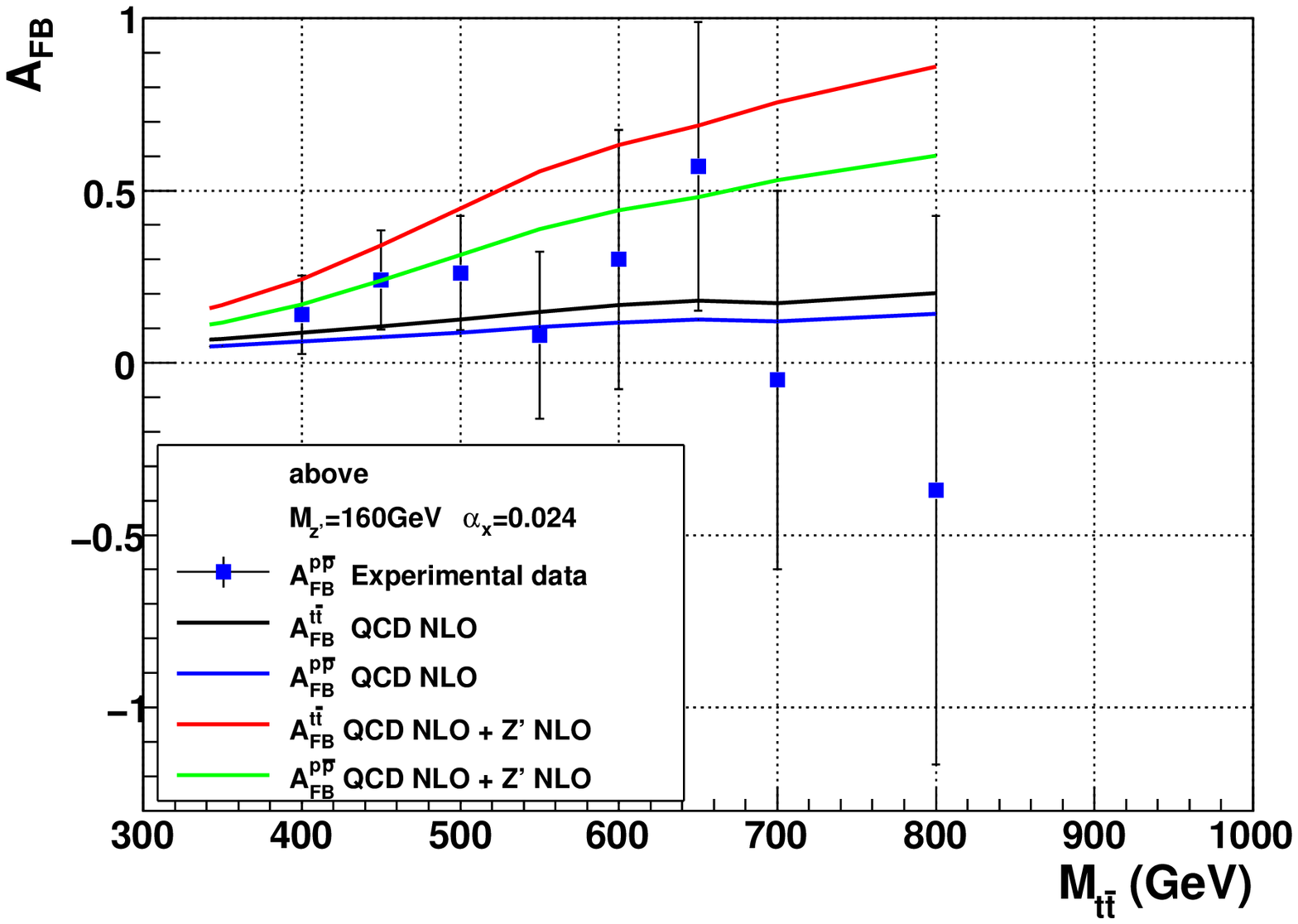}
\includegraphics[width=0.5\textwidth]
{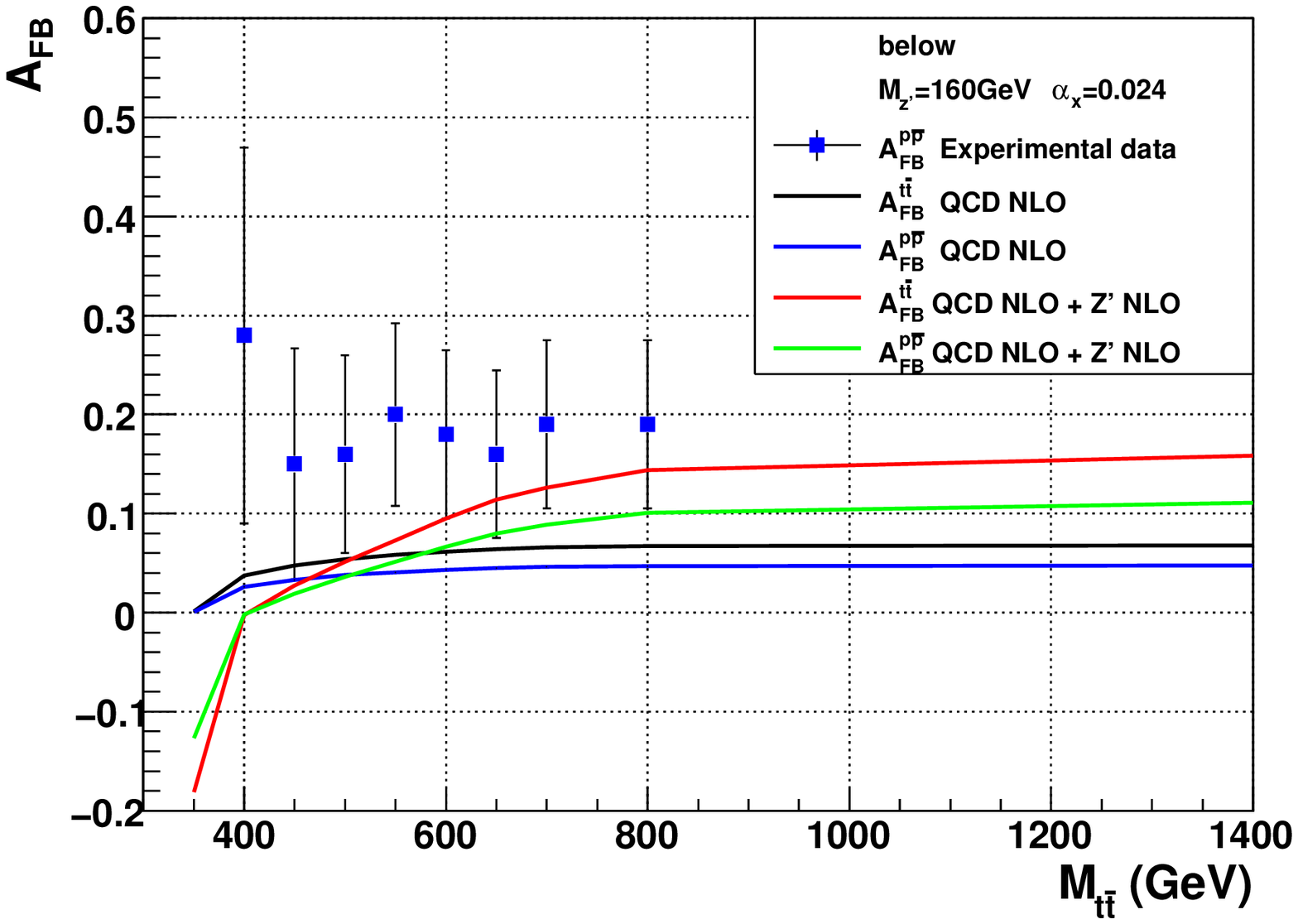} }}
 \caption{\label{asymmetry}
 Forward-backward asymmetry distributions
  as a function of $M_{t\bar{t}}$ with $\mu_r=\mu_f=m_t$.  }
\end{figure}

Histograms of the forward backward asymmetry $A_{FB}^{t\bar{t}}$ as
a function of $M_{t\bar{t}}$ in $t\bar{t}$ rest frame are drawn in
Fig.~\ref{asymmetry}, where $A_{FB}^{t\bar{t}}$ in each bin is
defined as
\begin{equation}
A_{FB}^{t\bar{t}}=\frac{N(\Delta Y>0)-N(\Delta Y<0)}{N(\Delta
Y>0)+N(\Delta Y<0)}
\end{equation}
where $\Delta Y=Y_t-Y_{\bar{t}}$ denotes the difference between the
$t$ and $\bar{t}$ rapidities. The total $A_{FB}^{t\bar{t}}$ is
calculated to be $15.8\%$, by summing $A_{FB}^{t\bar{t}}$ in each
bin multiplied by their corresponding weights. To compare directly
with experimental data \cite{CDFnote9724}, we also draw the so
called ``above" and ``below" $A_{FB}$ distribution at the bottom
 of Fig.~\ref{asymmetry} in which $A_{FB}$ is measured or calculated for
$M_{t\bar{t}}$ above or below a certain value. It should be noted
that the experimental data are measured in the $p\bar{p}$ lab frame.
As the simplest approximation, we utilize the relation
$A_{FB}^{p\bar{p}}\approx0.7A_{FB}^{t\bar{t}}$. Obviously more
measurements are needed to decrease the experimental uncertainties
in order to confirm/exclude the FVZM.

Total cross sections and total $A_{FB}^{t\bar{t}}$ and are shown
together in Table~\ref{Afbcs}.  For total cross section, the
$Z'$-born contribution decreases the NLO QCD cross section.
Including the $Z'$-NLO corrections makes the cross section even
smaller although these corrections are not significant. On the
contrary $A_{FB}^{t\bar{t}}$ is sensitive to $Z'$-NLO correction and
can drop about $30\%$ from the $Z'$-Born value.

\begin{table}[htb]
\caption{\label{Afbcs} $A_{FB}^{t\bar{t}}$ and total cross sections
with $m_{Z'}=160\mbox{GeV}, \alpha_X=0.024, \mu_r=\mu_f=m_t.$ }
\center \small{
\begin{tabular}{cccc}\hline\hline
&QCD NLO\   &\ QCD NLO+$Z'$ Born\ &\ QCD NLO+$Z'$ NLO\\\hline
$A_{FB}^{t\bar{t}}(\%)$&6.8 &22.2&15.8\\
Total cross section(pb) &6.29 &5.52&5.13\\
\hline\hline
\end{tabular}
}
\end{table}

\section{Conclusions and discussions \label{summary}}

In this paper, we calculate the top quark differential cross section
and asymmetry up to $\mathscr{O}(\alpha_s^2\alpha_X)$ in a flavor
violating $Z'$ model (FVZM). In the FVZM, the leading $Z'$
contribution can induce the measured top asymmetry, while the
differential distribution of $M_{t\bar t}$ does not fit measurement
well. After including the higher order contribution, the
differential distribution can be improved while the top asymmetry is
still in agreement with the observed value.

QCD soft gluon resummation effects for the top quark pair production
in the SM have been considered in Ref.~\cite{Almeida:2008ug}. Such
effects do not change the $M_{t\bar{t}}$ distribution significantly.
It is expected that resummation effect in the FVZM is similar to
that in the SM because the internal $Z'$ contributions has nothing
to do with the soft gluon radiations from external quark legs. Such
effects are under investigation~\cite{WXZ}.

\section*{Acknowledgment}
This work was supported in part by the Natural Sciences Foundation
of China (Nos. 10775001, 10635030).

\appendix

\section*{Appendix A. Renormalization Constants}

Renormalization constants are needed when calculating
Fig.~\ref{zpvirtualdiagram} and Fig.~\ref{pureQCDvirtualdiagram}.
$\delta Z_u$, $\delta Z_t$, $\delta Z_A$ corresponds to up quark,
top quark and gluon on-mass-shell renormalization constants
respectively.
 $\delta Z_g$ is the coupling renormalization
constants.

\begin{equation}
\begin{array}{rl}
\delta Z_u =&
\frac{\alpha_s}{2\pi}\frac{(4\pi)^\epsilon}{\Gamma(1-\epsilon)}
(-\frac{1}{2}
C_F\frac{1}{\epsilon_{UV}}+\frac{1}{2} C_F\frac{1}{\epsilon_{IR}}),\\\\
\delta Z_t =&
\frac{\alpha_s}{2\pi}\frac{(4\pi)^\epsilon}{\Gamma(1-\epsilon)}
(-\frac{1}{2}
C_F\frac{1}{\epsilon_{UV}}-C_F\frac{1}{\epsilon_{IR}}-2
C_F+\frac{3}{2}
 C_F \log(\frac{m_t^2}{\mu^2})), \\\\
\delta Z_A =&
\frac{\alpha_s}{2\pi}\frac{(4\pi)^\epsilon}{\Gamma(1-\epsilon)}
((\frac{5}{6} C_A-\frac{2}{3} T_F n_{lf}-\frac{2}{3} T_F
n_{hf})\frac{1}{\epsilon_{UV}} -(\frac{5}{6} C_A-\frac{2}{3} T_F
n_{lf})\frac{1}{\epsilon_{IR}}\\\\
&+\frac{2}{3}
T_F(\log(\frac{m_c^2}{\mu^2})+\log(\frac{m_d^2}{\mu^2})+\log(\frac{m_t^2}{\mu^2}))),\\\\
 \delta Z_g =& \frac{\alpha_s}{2\pi}
\frac{(4\pi)^\epsilon}{\Gamma(1-\epsilon)}((\frac{1}{\epsilon_{UV}}-\log(\frac{\mu_r^2}{\mu^2}))(-\frac{11}{12}
C_A+\frac{1}{3} T_F +\frac{1}{3} T_F n_{hf})),
\end{array}
\end{equation}
where $C_F=4/3$, $C_A=3$, $T_F=1/2$, $n_{lf}=3$ is the number of
light quark flavors, $n_{hf}=3$ is the number of heavy quark
flavors, $\mu$ is the energy scale introduced in dimensional
regularization, $\mu_r$ is the renormalization scale.

\begin{figure}[htbp]
\centerline{\hbox{
\includegraphics[height=2.5cm,width=2.5cm]
{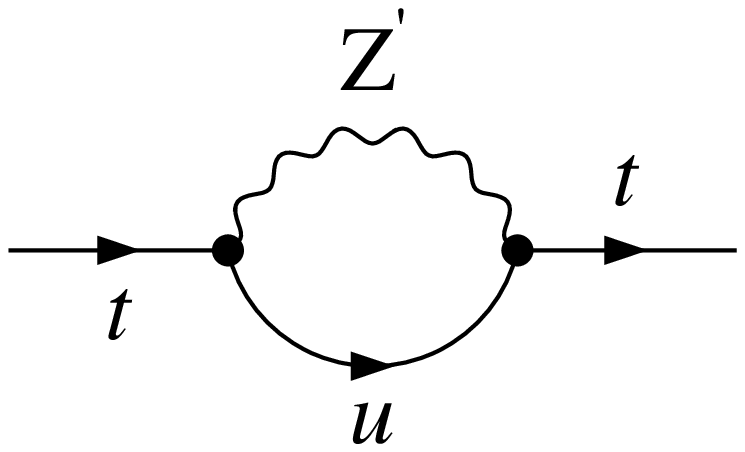}
\includegraphics[height=2.5cm,width=2.5cm]
{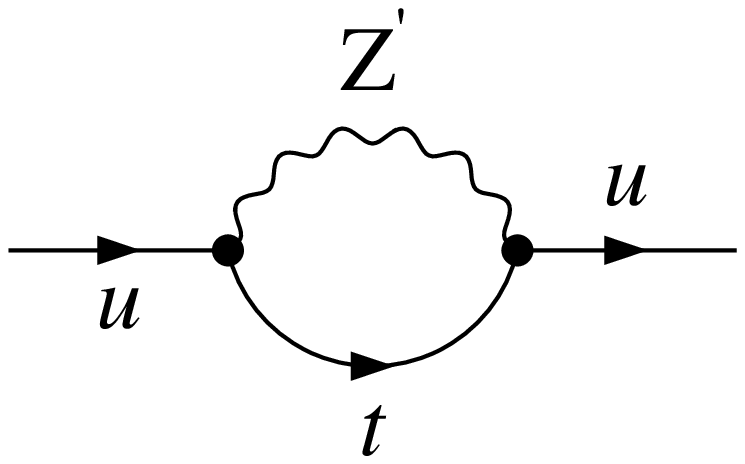}
 }}
\caption{\label{tuselfenergy}Self energy diagrams for counterterm
calculation.}
\end{figure}

There are new contributions to top and up quark field
renormalization constants, when we calculate
Fig.~\ref{zpvirtualdiagram}.
 Top quark field renormalization constants $\delta
Z_t^{V}$, $\delta Z_t^A$ are calculated from $Z'$ induced top
Self-energy $-i\Sigma_{t}(p\!\!/)$\cite{Beenakker:1993yr}, as showed
in Fig.~\ref{tuselfenergy}.

\begin{equation}
\begin{array}{rl}
\Sigma_t(p\!\!/) =&
p\!\!/(\Sigma_t^V(p^2)+\Sigma_t^A(p^2)\gamma^5)+m_t\Sigma_t^S(p^2),\\\\
 \delta Z_t^V =&
\Sigma_t^V(p^2)|_{p^2=m_t^2}+2m_t^2\frac{\partial}{\partial
p^2}(\Sigma_t^V(p^2)|_{p^2=m_t^2}+ \Sigma_t^S(p^2)|_{p^2=m_t^2}),\\\\
\delta Z_t^A =& \Sigma_t^A(p^2)|_{p^2=m_t^2}.
\end{array}
\end{equation}

The counterterm for $t\bar t g$ vertex  is written as
\begin{equation}
(-i g_s T^a \gamma^\rho)(\delta Z_t^V + \delta Z_t^A \gamma^5),
\end{equation}
where the vector and axial vector parts are
\begin{equation}
\begin{array}{rl}
\delta Z_t^V =& -{\frac{(4\pi)^\epsilon}{\Gamma(1-\epsilon)}}
\frac{{g_X}^2 }{32 m_t
   ^2 \pi ^2}\\\\
   &\{\frac{m_t ^2}{\epsilon_{{UV}}}+[(2 ({m_t}^2+m_u
   ^2-{m_{z'}}^2) \frac{\partial}{\partial p^2}
   {B_0}(p^2,{m_u}^2,{m_{z'}}^2)|_{p^2=m_t^2}-1) m_t
   ^2\\\\
   &+({m_u}^2-{m_{z'}}^2)
   B_0(0,{m_u}^2,{m_{z'}}^2)+(-{m_u}^2+m_t ^2+{m_{z'}}^2)
   B_0(m_t ^2,{m_u}^2,{m_{z'}}^2)]\},\\\\
\delta Z_t^A =& -{\frac{(4\pi)^\epsilon}{\Gamma(1-\epsilon)}}
\frac{{g_X}^2}{32{m_t}^2 \pi ^2}\\\\
&\{\frac{{m_t}^2}{ \epsilon_{{UV}}}+[-{m_t}^2+({m_{z'}}^2-m_u
   ^2) B_0(0,m_u ^2,{m_{z'}}^2)\\\\
   &+({m_t}^2+m_u ^2-{m_{z'}}^2)
   B_0({m_t}^2,m_u ^2,{m_{z'}}^2)]\}.
\end{array}
\end{equation}

Field renormalization constants of the up quark can be calculated
similarly and the counterterm for $u\bar u g$ vertex  is written as
\begin{equation}
(-i g_s T^a \gamma^\rho)(\delta Z_u^V + \delta Z_u^A \gamma^5),
\end{equation}
where
\begin{equation}
\begin{array}{rl}
\delta Z_u^V =& -{\frac{(4\pi)^\epsilon}{\Gamma(1-\epsilon)}}
\frac{{g_X}^2 }{32 m_u
   ^2 \pi ^2}\\\\
   &\{\frac{m_u ^2}{\epsilon_{{UV}}}+[(2 ({m_t}^2+m_u
   ^2-{m_{z'}}^2) \frac{\partial}{\partial p^2}
   {B_0}(p^2,{m_t}^2,{m_{z'}}^2)|_{p^2=m_u^2}-1) m_u
   ^2\\\\
   &+({m_t}^2-{m_{z'}}^2)
   B_0(0,{m_t}^2,{m_{z'}}^2)+(-{m_t}^2+m_u ^2+{m_{z'}}^2)
   B_0(m_u ^2,{m_t}^2,{m_{z'}}^2)]\},\\\\
\delta Z_u^A =& -{\frac{(4\pi)^\epsilon}{\Gamma(1-\epsilon)}}
\frac{{g_X}^2}{32 m_u
   ^2 \pi ^2}\\\\
   & \{\frac{m_u ^2}{\epsilon_{{UV}}}+[-m_u
   ^2+({m_{z'}}^2-{m_t}^2)
   B_0(0,{m_t}^2,{m_{z'}}^2)\\\\
   &+({m_t}^2+m_u ^2-{m_{z'}}^2)
   B_0(m_u ^2,{m_t}^2,{m_{z'}}^2)]\}.
\end{array}
\end{equation}

\section*{Appendix B.  Soft part of the real gluon emission cross section}

Soft real squared amplitude $|{\cal{M}}_{q\bar q\to t\bar t
g}^{soft}|^2$, which is expressed as
$\left[2\mathscr{R}\left(f^{r*}_s
f^r_X\right)\alpha^2_s\alpha_X\right]_{\text{Soft}}$ in
Eq.~[\ref{realsoft}], can be obtained by the interference of
diagrams in Figs.~\ref{qcdrealgluonradiation} and
\ref{zprealgluonradiation} with requirement that the gluon's energy
is smaller than $\delta_s \sqrt{s}/2$ . $|{\cal{M}}_{q\bar q\to
t\bar t g}^{soft}|^2$ can be expressed as
\begin{equation}
|{\cal M}_{q\bar q\to t\bar t g}^{soft}|^2=|{\cal M}_{q\bar q\to
t\bar t }|^2\sum\limits_{i,j=1}^4\frac{C_{ij}}{C_0}S_{ij},
\end{equation}
where $|{\cal M}_{q\bar q\to t\bar t}|^2=2\mathscr{R}\left( f_s^*
f_X \right)^D \alpha_s \alpha_X$ is the interference term of the two
born diagrams in Fig.~\ref{tree}. $C_{ij}=C_{ji}$ is the color
factor of interference terms with one gluon emitting from external
leg $i$ of one diagram and from external leg $j$ of another diagram.
\begin{equation}
\begin{array}{l}
C_{12} = C_{14} = C_{23} = C_{34} = -C_F/2,\\\\
C_{11} = C_{22} = C_{13} = C_{24} = C_{33} = C_{44} = C_A C_F^2.
\end{array}
\end{equation}
$C_0=C_A C_F$ is the color factor of the interference of the two
diagrams in Fig.~\ref{tree}. $S_{ij}=S_{ji}$ are the soft factors of
the corresponding interference terms. They are calculated by using
eikonal approximation method. According to
Ref.~\cite{Harris:2001sx},
\begin{equation}
\begin{array}{rl}
S_{ij}=&\frac{\alpha_s}{2\pi}\frac{(4\pi)^\epsilon}{\Gamma(1-\epsilon)}
(\frac{\mu^2}{s})^\epsilon\\\\
&\times\frac{1}{\pi}(\frac{4}{s})^{-\epsilon}\int_0^{\delta_s\frac{\sqrt{s}}{2}}dE_q\int_0^\pi
d\theta_1 \int_0^\pi
d\theta_2\left(\eta_i\eta_j\frac{p_i^\mu}{p_i\cdot
q}\frac{p_j^\nu}{p_j\cdot q}(-g^{\mu\nu})\right)
E_q^{1-2\epsilon}\sin^{1-2\epsilon}\theta_1\sin^{-2\epsilon}\theta_2,
\end{array}
\end{equation}
where $\eta_{i}$ is a sign which is positive for outgoing fermion or
incoming antifermion, and is negative for incoming fermion or
outgoing antifermion.

\begin{equation}
\begin{array}{rl}
S_{11}=S_{22}=&0,\\\\
 S_{12} =& \frac{\alpha_s}{2
\pi}{\frac{(4\pi)^\epsilon}{\Gamma(1-\epsilon)} }
 (\frac{1}{\epsilon
   _{{IR}}^2}-2 \log (\frac{\sqrt{s}
   {\delta s}}{\mu })\frac{1}{\epsilon
   _{{IR}}}+2 \log
   ^2(\frac{\sqrt{s} {\delta s}}{\mu
   })-\frac{\pi ^2}{6}),\\\\
S_{13} = S_{24} =& \frac{\alpha_s}{2
\pi}{\frac{(4\pi)^\epsilon}{\Gamma(1-\epsilon)} }
   \{\frac{1}{2}\frac{1}{
   \epsilon _{{IR}}^2}-[\log (\frac{\sqrt{s}
   {\delta s}}{\mu })+\frac{1}{2} \log
   (\frac{(1-\beta  \cos \theta )^2}{1-\beta
   ^2})]\frac{1}{\epsilon _{IR}}\\\\
   &-\frac{1}{4} \log ^2(\frac{\beta
   +1}{1-\beta })+\log ^2(\frac{\sqrt{s}
   {\delta s}}{\mu })+\frac{1}{2} \log
   ^2(\frac{1-\beta }{1-\beta  \cos \theta
   })\\\\
   &+\log (\frac{\sqrt{s} {\delta
   s}}{\mu }) \log (\frac{(1-\beta  \cos
   \theta )^2}{1-\beta
   ^2})+{li}_2(\frac{\beta  (\cos
   \theta -1)}{1-\beta
   })-{li}_2(-\frac{\beta  (\cos
   \theta +1)}{1-\beta  \cos \theta
   })-\frac{\pi ^2}{12}\},\\\\
S_{14} = S_{23} =& \frac{\alpha_s}{2
\pi}{\frac{(4\pi)^\epsilon}{\Gamma(1-\epsilon)} }
 \{-\frac{1}{2}
   \frac{1}{\epsilon _{{IR}}^2}+[\log (\frac{\sqrt{s}
   {\delta s}}{\mu })+\frac{1}{2} \log
   (\frac{(\beta  \cos \theta +1)^2}{1-\beta
   ^2})]\frac{1}{\epsilon _{{IR}}}\\\\
   &+\frac{1}{4}
   \log ^2(\frac{\beta +1}{1-\beta })-\log
   ^2(\frac{\sqrt{s} {\delta s}}{\mu
   })-\frac{1}{2} \log ^2(\frac{1-\beta
   }{\beta  \cos \theta +1})\\\\
   &-\log
   (\frac{\sqrt{s} {\delta s}}{\mu })
   \log (\frac{(\beta  \cos \theta
   +1)^2}{1-\beta
   ^2})-{li}_2(\frac{\beta  (-\cos
   \theta -1)}{1-\beta
   })+{li}_2(-\frac{\beta  (1-\cos
   \theta )}{\beta  \cos \theta
   +1})+\frac{\pi ^2}{12}\},\\\\
S_{33}= S_{44}=& \frac{\alpha_s}{2
\pi}{\frac{(4\pi)^\epsilon}{\Gamma(1-\epsilon)} }
 (\frac{1}{\epsilon _{{IR}}}+\frac{1}{\beta
}\log
   (\frac{\beta +1}{1-\beta })-2
   \log (\frac{\sqrt{s} {\delta s}}{\mu
   })),\\\\
S_{34} =& \frac{\alpha_s}{2
\pi}{\frac{(4\pi)^\epsilon}{\Gamma(1-\epsilon)} }
   \{-\frac{(\beta ^2+1)}{2 \beta} \log
   (\frac{\beta +1}{1-\beta })\frac{1}{\epsilon _{IR}}\\\\
   &-\frac{(\beta ^2+1)}{\beta }
   [\frac{1}{4} \log ^2(\frac{\beta
   +1}{1-\beta })-\log (\frac{\sqrt{s}
   {\delta s}}{\mu }) \log
   (\frac{\beta +1}{1-\beta
   })+{li}_2(\frac{2 \beta }{\beta
   +1})]\},
\end{array}
\end{equation}
in which $\beta=\sqrt{1-4m_t^2/s}$ and $\theta$ is the angle between
the incoming $u$ and outgoing $t$ quark in $t\bar t$ rest frame.


\begin{thebibliography}{10}

\bibitem{PhysRevLett.100.142002}
V.~M. Abazov {\em et~al.},
\newblock Phys. Rev. Lett. {\bf 100}, 142002 (2008), arXiv:0712.0851 [hep-ex].

\bibitem{Aaltonen:2008hc}
CDF, T.~Aaltonen {\em et~al.},
\newblock Phys. Rev. Lett. {\bf 101}, 202001 (2008), arXiv:0806.2472 [hep-ex].

\bibitem{Aaltonen:2009iz}
CDF, T.~Aaltonen {\em et~al.},
\newblock Phys. Rev. Lett. {\bf 102}, 222003 (2009), arXiv:0903.2850 [hep-ex].

\bibitem{Kuhn:1998jr}
J.~H. Kuhn and G.~Rodrigo,
\newblock Phys. Rev. Lett. {\bf 81}, 49 (1998), arXiv:hep-ph/9802268.

\bibitem{Kuhn:1998kw}
J.~H. Kuhn and G.~Rodrigo,
\newblock Phys. Rev. {\bf D59}, 054017 (1999), arXiv:hep-ph/9807420.

\bibitem{Antunano:2007da}
O.~Antunano, J.~H. Kuhn, and G.~Rodrigo,
\newblock Phys. Rev. {\bf D77}, 014003 (2008), arXiv:0709.1652 [hep-ph].

\bibitem{Almeida:2008ug}
L.~G. Almeida, G.~Sterman, and W.~Vogelsang,
\newblock Phys. Rev. {\bf D78}, 014008 (2008), arXiv:0805.1885 [hep-ph].

\bibitem{Frampton:2009rk}
P.~H. Frampton, J.~Shu, and K.~Wang,
\newblock Phys. Lett. {\bf B683}, 294 (2010), arXiv:0911.2955 [hep-ph].

\bibitem{Shu:2009xf}
J.~Shu, T.~M.~P. Tait, and K.~Wang,
\newblock Phys. Rev. {\bf D81}, 034012 (2010), arXiv:0911.3237 [hep-ph].

\bibitem{Jung:2009jz}
S.~Jung, H.~Murayama, A.~Pierce, and J.~D. Wells,
\newblock Phys. Rev. {\bf D81}, 015004 (2010), arXiv:0907.4112 [hep-ph].

\bibitem{Cheung:2009ch}
K.~Cheung, W.-Y. Keung, and T.-C. Yuan,
\newblock Phys. Lett. {\bf B682}, 287 (2009), arXiv:0908.2589 [hep-ph].

\bibitem{Cao:2010zb}
Q.-H. Cao, D.~McKeen, J.~L. Rosner, G.~Shaughnessy, and C.~E.~M.
Wagner,
\newblock (2010), arXiv:1003.3461 [hep-ph].

\bibitem{Djouadi:2009nb}
A.~Djouadi, G.~Moreau, F.~Richard, and R.~K. Singh,
\newblock (2009), arXiv:0906.0604 [hep-ph].

\bibitem{Jung:2009pi}
D.-W. Jung, P.~Ko, J.~S. Lee, and S.-h. Nam,
\newblock (2009), arXiv:0912.1105 [hep-ph].

\bibitem{Cao:2009uz}
J.~Cao, Z.~Heng, L.~Wu, and J.~M. Yang,
\newblock Phys. Rev. {\bf D81}, 014016 (2010), arXiv:0912.1447 [hep-ph].

\bibitem{Barger:2010mw}
V.~Barger, W.-Y. Keung, and C.-T. Yu,
\newblock (2010), arXiv:1002.1048 [hep-ph].

\bibitem{Arhrib:2009hu}
A.~Arhrib, R.~Benbrik, and C.-H. Chen,
\newblock Phys. Rev. {\bf D82}, 034034 (2010), arXiv:0911.4875 [hep-ph].

\bibitem{Mertig:1990an}
R.~Mertig, M.~Bohm, and A.~Denner,
\newblock Comput. Phys. Commun. {\bf 64}, 345 (1991).

\bibitem{Hahn:1998yk}
T.~Hahn and M.~Perez-Victoria,
\newblock Comput. Phys. Commun. {\bf 118}, 153 (1999), arXiv:hep-ph/9807565.

\bibitem{Ellis:2007qk}
R.~K. Ellis and G.~Zanderighi,
\newblock JHEP {\bf 02}, 002 (2008), arXiv:0712.1851 [hep-ph].

\bibitem{Harris:2001sx}
B.~W. Harris and J.~F. Owens,
\newblock Phys. Rev. {\bf D65}, 094032 (2002), arXiv:hep-ph/0102128.

\bibitem{CDFnote9724}
M.~Tecchio {\em et~al.},
\newblock Measurement of the dependence of the forward-backward asymmetry in
  top pair production on mtt,
\newblock CDFnote 9724.

\bibitem{WXZ}
Y.-k. Wang, B.~Xiao, and S.-h. Zhu,
\newblock (unpublished).

\bibitem{Beenakker:1993yr}
W.~Beenakker,{\em et~al.},
\newblock Nucl. Phys. {\bf B411}, 343 (1994).

\end{thebibliography}
\end{document}